\documentclass[12pt, preprint]{aastex}
\usepackage[dvips]{color}

\def\ltsima{$\;\buildrel < \over \sim \;$}
\def\simlt{\lower.5ex \hbox{\ltsima}}
\def\gtsima{$\;\buildrel > \over \sim \;$}
\def\simgt{\lower.5ex \hbox{\gtsima}}

\newcommand{\re}{\textcolor[rgb]{0,0,0}}

\shorttitle{Spitzer spectral line mapping of protostellar outflows}
\shortauthors{Neufeld et al.}

\begin{document}

\title{Spitzer spectral line mapping of protostellar outflows\\
 I.\ Basic data and outflow energetics}
\author{David A.~Neufeld\altaffilmark{1}, Brunella Nisini\altaffilmark{2}, Teresa Giannini\altaffilmark{2}, Gary J.~Melnick\altaffilmark{3}, Edwin A.\ Bergin\altaffilmark{4}, Yuan Yuan\altaffilmark{1}, S\'ebastien Maret\altaffilmark{5}, Volker Tolls\altaffilmark{3}, Rolf G\"usten\altaffilmark{6}, and Michael J.\ Kaufman\altaffilmark{7}}

\altaffiltext{1}{Department of Physics and Astronomy, Johns Hopkins University,
3400 North Charles Street, Baltimore, MD 21218, USA}
\altaffiltext{2}{INAF -- Osservatorio Astronomico di Roma, via Frascati 33, I-00040 Monte Porzio, Italy}
\altaffiltext{3}{Harvard-Smithsonian Center for Astrophysics, 60 Garden Street, 
Cambridge, MA 02138, USA}
\altaffiltext{4}{Department of Astronomy, University of Michigan, 825 Dennison Building, Ann Arbor, MI 48109, USA}
\altaffiltext{5}{Laboratoire d'Astrophysique de Grenoble, Observatoire de Grenoble, Universit\'e Joseph
Fourier, CNRS, UMR 571, BP 53, F-38041 Grenoble, France}
\altaffiltext{6}{Max Planck Institut f\"ur Radioastronomie, Auf dem H\"ugel 69, D-53121, Bonn, Germany}
\altaffiltext{7}{Department of Physics and Astronomy, San Jose State University, One Washington Square, San Jose, CA 95192, USA}

\begin{abstract}

We report the results of spectroscopic mapping observations carried out toward protostellar outflows in the BHR71, L1157, L1448, NGC 2071, and VLA 1623 molecular regions using the Infrared Spectrograph (IRS) of the {\it Spitzer Space Telescope}.  These observations, covering the $5.2 - 37\rm\,\mu m$ spectral region, provide detailed maps of the 8 lowest pure rotational lines of molecular hydrogen and of the [SI] 25.25$\rm\,\mu m$  and [FeII] 26.0$\rm\,\mu m$ fine structure lines.  The molecular hydrogen lines, believed to account for a large fraction of the radiative cooling from warm molecular gas that has been heated by a non-dissociative shock, allow the energetics of the outflows to be elucidated.  Within the regions mapped towards these 5 outflow sources, total H$_2$ luminosities ranging from 0.02 to 0.75 L$_{\odot}$ were inferred for the sum of the 8 lowest pure rotational transitions. By contrast, the much weaker [FeII] 26.0$\rm\,\mu m$ fine structure transition traces faster, dissociative shocks; here, only a small fraction of the fast shock luminosity emerges as line radiation that can be detected with {\it Spitzer}/IRS.

\end{abstract}


\section{Introduction}

It has long been recognized that the formation of stars in molecular clouds is often accompanied by supersonic outflows (e.g. Bally \& Lada 1983).  Where protostellar outflows strike nearby molecular material, shock waves can be created; these compress and heat the gas, and can modify its composition.  Interstellar shocks have been the subject of extensive study, both observational and theoretical (summarized, for example, in the review of Draine \& McKee 1993).  Slow shocks -- propagating at speeds less than $\sim 50 \rm  \, km\, s^{-1}$ (in clouds with typical magnetic fields) -- leave the gas predominantly molecular in their wake, although the chemical composition can be modified by two processes: (1) the sputtering of dust grains and their icy mantles, which releases material into the gas phase; and (2) the enhancement of gas-phase chemical reactions that possess activation energy barriers; these may be negligibly slow at the typical temperatures of molecular clouds but rapid at the elevated temperatures behind shock waves.  Faster shocks can lead to the collisional dissociation of molecules and the ionization of the resultant atoms.

\re{ The Infrared Spectrograph (IRS) of the {\it Spitzer Space Telescope} provides a valuable tool for observing emissions from shocked interstellar gas.  With its coverage of $5.2 - 37\rm\,\mu m$ spectral region, IRS has provided access to (1) the lowest 8 pure rotational transitions of molecular hydrogen (e.g.\ Neufeld et al. 2006a, hereafter N06) -- the dominant constituent of molecular clouds; (2) pure rotational transitions of HD (e.g.\ Neufeld et al. 2006b), H$_2$O, and OH (e.g. Melnick et al.\ 2008); (3) rovibrational transitions of CO$_2$ and C$_2$H$_2$ (e.g. Sonnentrucker et al.\ 2006; Sonnentrucker, Gonz\'alez-Alfonso \& Neufeld 2007); (4) fine structure emissions of several atoms and atomic ions (Neufeld et al.\ 2007, hereafter N07).   Observations of shocked material at mid-infrared wavelengths offer significant advantages over ultraviolet, optical or even near-infrared studies: they are affected less by interstellar extinction, and they can probe cooler material than can be observed at shorter wavelengths.}
   
In this paper, we present the results of spectroscopic mapping observations carried out using the {\it Spitzer}/IRS) toward shocked material associated with five protostellar outflows: BHR71, L1157, L1448, NGC 2071, and VLA 1623.  In \S 2 below, we discuss the properties of the individual sources and the rationale underlying our source selection.  In \S 3, we describe the observations, discuss our data reduction procedure, and present average spectra and spectral line maps for each source.  In \S 4, we discuss the basic features of the H$_2$ rotational emissions and atomic fine structure emissions that we detected; here we analyse the overall energetics for each source, describe the correlations among the various detected line emissions, 
and compare our H$_2$ spectral line maps with continuum maps obtained previously using {\it Spitzer}'s Infrared Array Camera (IRAC) and extracted by us from the {\it Spitzer} archive.  

\section{Source selection}

The sample selected for the present investigation is composed of five well-studied
protostellar outflows, namely L1448, BHR71, VLA 1623, L1157
and NGC 2071,  that are known to contain warm molecular gas from ground-based observations of H$_2$ vibrational emissions.

\re{ Our sample is not intended to provide an unbiased survey of protostellar outflows.}  The first four sources listed above are 
\re{ all}  outflows driven by low-mass, low luminosity (L $<$ 70 L$_\odot$) protostars that provide a particularly ''clean'' laboratory for studying shock physics. The fifth source, NGC 2071, contains a cluster of intermediate-mass protostars with a higher luminosity ($L \sim 500\,L_\odot$) and thus 
allows us to contrast the outflows associated with low-mass objects 
with those driven by intermediate mass protostars. 
The new observations of this source complement previous Spitzer observations
of a single $1^\prime \times 1^\prime$ region that have led recently to the first
detection of spatially-extended water (Melnick et al.\ 2008).

All the sources in our target list have been extensively studied at several 
wavelengths using a variety of different facilities; and all of them 
will be observed as part of the Herschel
Guaranteed Time Key Program “Water in Starforming Regions with Herschel” (WISH).  
\re{ We have provided a description of NGC 2071 in a recent paper (Melnick et al.\ 2008); the properties of the other four sources are reviewed below.}

\subsection{BHR71}

BHR 71 (also known as DC 297.2--2.8 or Sa 136) is a well-isolated 
Bok globule located at a distance of $\sim\,$200 pc toward the constellation Musca 
(Sandqvist 1977; Bourke, Hyland, \& Robinson 1995a; Bourke et al. 1995b).  CO observations of this
source reveal a highly collimated bipolar outflow 
(Bourke et al. 1997) associated with IRAS 11590--6452,
which was classified as a very young Class 0
protostar with a luminosity of $\sim\,$9 L$_{\odot}$ .
Higher spatial resolution observations using ISOCAM, combined with
ground-based near-infrared $K^{\prime}$-band (2.11$\:\mu$m) imaging and 
$^{12}$CO $J = 2 \rightarrow\ 1$ observations, show that the IRAS source
is likely composed of two embedded protostars, IRS~1 and IRS~2, with a projected
separation of $\sim\,$17$^{\prime\prime}$ (3400 AU; Bourke 2001).  IRS~1 and
IRS~2 each drive a molecular outflow, though the significantly larger outflow, and 
the one of greater interest here, is driven by IRS~1.

Recent observations of the circumstellar envelope around IRS~1 and IRS~2 by 
Chen et al.~(2008) using both the Australian
Telescope Compact Array and {\em Spitzer}, reveal masses of 2.12 and 0.05 $M_{\odot}$,
hydrogen (H+2H$_2$) densities of 2.6 and 2.2$\times$10$^7$ cm$^{-3}$,
and hydrogen column densities of 9.4 and 2.6$\times$10$^{23}$ cm$^{-2}$, respectively.  
Fits to the spectral energy distribution of IRS~1 and
IRS~2 based on these more recent data imply bolometric luminosities for these
objects of 13.5 and 0.5 $L_{\odot}$, respectively.

The {\em Spitzer}/IRAC image of BHR 71 is shown in Figure 1.
Larger-scale, multi-transition CO maps reveal that the BHR~71 outflow extends 
$\pm\,$4$^{\prime}$ from IRS~1.  Excitation analysis implies
30$\;< T <\,$50$\;$K and 10$^5\,<\,n($H$_2)\,<\,$3$\times$10$^5$ cm$^{-3}$
for the IRS~1 outflow.  Though smaller, the outflow from IRS~2 appears to
contain warmer gas, i.e., $T >\,$300~K (Parise et al. 2006).  
The CO column densities associated with
the outflow lobes from IRS~1 are estimated to be 2--3$\times$10$^{17}$ cm$^{-2}$.
Assuming $n({\rm CO})/n({\rm H}_2)\,=\,$10$^{-4}$, the inferred H$_2$ column density
toward each lobe is 2--3$\times$10$^{21}$ cm$^{-2}$.
Bourke et al. (1997) find the
observed morphology and velocity structure of the flow are well accounted for with 
a simple model of a biconical outflow with a semi-opening angle of 15$^{\rm o}$, 
in which the gas moves outward with a constant radial velocity (with
respect to the cone apex) of $\sim\,$28 km s$^{-1}$ and that is viewed
nearly perpendicular to its symmetry axis. The inclination
of the outflow axis from the line of sight is found to be $\sim\,$84$^{\rm o}$.
The flow
masses within each lobe, taking into account the mass in the velocity range of the
ambient cloud and optical depth effects of the flowing gas, are 0.3 and 1.0 $M_{\odot}$
for the southern and northern lobes, respectively. The mechanical luminosity of the
molecular outflow, derived from $^{12}$CO (1--0) and $^{12}$CO (1--0) observations,
is $\sim\,$0.5 $L_{\odot}$ and its dynamical age is $\sim\,$10$^4$ yr
(Bourke et al. 1997).

\subsection{L1157}

The well-collimated molecular outflow in L1157 lies deeply embedded within an 
extended circumstellar envelope with which it is interacting (Gueth et al. 2003; 
Beltr\'{a}n et al. 2004).
The driving source is believed to be a Class 0, low-luminosity (11 $L_{\odot}$)
IRAS source, IRAS 20386+6751 (Andre et al. 1993).  The outflow is oriented
approximately north-south and extends over about 5$^{\prime}$.
The distance to L1157 is assumed to share that of other clouds in Cepheus.
Unfortunately, estimates for the distances of these clouds cover a fairly wide range: 200 to 450 pc
(e.g.\ Kun 1998, 
\re{ Looney et al.\ 2007}).  Though the distance appropriate for L1157 is uncertain and
remains a matter of debate, we adopt a value of 440~pc (Viotti 1969).

Figure 2 shows the {\em Spitzer}/IRAC image of L1157.  There are notable
asymmetries between the northern and southern lobe, but the morphology is clearly
S-shaped indicating the presence of an underlying precessing jet.
Modeling of the outflow provides evidence for 3 or 4 separate episodes of mass
ejection from the driving source over a period of $\sim\,$15,000 years (Bachiller et al.
2001).
Observations of a number of excited molecular transitions clearly
indicate that shocks are present in both lobes.
Near-infrared H$_2$ v=1-0 S(1) + continuum imaging (Davis \& Eisl\"{o}ffel 1995)
reveal a number of knots of H$_2$ emission corresponding to the peaks seen in
the {\em Spitzer}/IRAC image, with the most prominent of these H$_2$ emission peaks
lying approximately 1.0$^{\prime}$ to the southeast and 
1.2$^{\prime}$ to the northwest of IRAS 20386+6751.  Similarly,
enhanced SiO and NH$_3$ (3,3) emission has been detected associated with the outflow 
(Mikami et al. 1992; Bachiller, Mart\'{i}n-Pintado, \& Fuente 1993), consistent
with elevated temperatures and an abundance increase tied to the passage of a shock.
Moreover, the spatial coincidence of
the SiO and NH$_3$ (3,3) emission with that of the H$_2$ v=1-0 S(1) emission
confirm that these emission lines arise in shocked material along the outflow.  

Multi-transition CO observations of the blueshifted southern lobe suggest that the
bulk of the outflow consists of moderately dense gas of $n$(H$_2)\,=\,$(1--3)$\times$10$^4$
cm$^{-3}$ heated to $T_{\rm kin} =\:$50--170~K (Hirano \& Taniguchi 2001).
The intense thermal emission lines of SiO and CS preferentially sample 
more compact regions within the CO-defined outflow lobe 
with densities, $n$(H$_2$), of 1--5$\times$10$^5$ cm$^{-3}$ 
(see Mikami et al. 1992; Nisini et al.\ 2007).  Bachiller \& P\'{e}rez Guti\'{e}rrez (1997) have obtained
estimates of the column density for a number of molecular species toward the driving
source and two positions in the southern lobe.  Using the reported column densities
for $^{13}$CO (1--0, 2--1) and assuming an abundance relative
to H$_2$ of 1.1$\times$10$^{-6}$ we obtain a total H$_2$ column density 
of 4.7$\times$10$^{21}$ cm$^{-2}$ toward the center of the outflow
($\alpha\,=\,$20$^{\rm h\:}$39$^{\rm m\:}$06.$^{\rm\!\!\! s}$19;
$\delta\,=\,+68^{\rm o\:}02^{\prime\:}15.^{\!\!\prime\prime}$9, J2000) and 
1.3$\times$10$^{22}$ cm$^{-2}$ $\Delta\alpha\,=\,20^{\prime\prime}$,
$\Delta\delta\,=\,-60^{\prime\prime}$ and
1.1$\times$10$^{22}$ cm$^{-2}$ $\Delta\alpha\,=\,35^{\prime\prime}$,
$\Delta\delta\,=\,-95^{\prime\prime}$ offset from the center position.  The
outflow mass and kinetic energy (not corrected for projection effects) 
derived from these CO observations are 
0.36 $M_{\odot}$ and 1.8$\times$10$^{44}$ erg, respectively, for the southern 
lobe and 0.26 $M_{\odot}$ and 1.9$\times$10$^{44}$ erg, respectively, for the 
northern lobe.

\subsection{L1448}

In the vicinity of Lynds 1448 there are a number of molecular outflows that have been detected via a variety
of observational techniques in the millimeter and infrared \citep[e.g., ][]{wc00}.    One of the more spectacular
outflows is associated with L1448-mm (L1448C) \citep{bac91}.  This flow, associated
with a Class 0 source, has been detected at a wide range of wavelengths and exhibits strong signatures of high velocity gas,
shock heating, and also shock chemistry.   The flow was first detected in CO emission with the highest velocities ($\sim 70$ km s$^{-1}$) displaying a high degree of collimation in emission morphology \citep{bac90}.   The high velocity CO gas is coincident with strong H$_2$ vibrational emission \citep{bll93}, with H$_2$ emission appearing illuminated on the leading edges of a bipolar cavity \citep{bac95}. 
The shocked gas is associated with an active chemistry with elevated abundances of SiO \citep{bmpf91}, CH$_3$OH \citep{j-s05}, and H$_2$O \citep{nisini99}.  
   
In Figure 3 we provide a finder chart illustrating our Spitzer IRS map coverage of the L1448-mm flow and also
delineate notable sources in the region. 
L1448-mm or L1448C is itself a binary system, as identified in {\it Spitzer} images, and the flow originates in the northern source, labelled as L1448C(N) in the nomenclature of \citet{jorg06}.    At the tip of the northern blue-shifted lobe lies another Class 0 protostar, L1448-IRS3 (L1448-N).  This source is also a binary \citep{looney00} with outflow emission that corresponds to the bright spots labelled as IRS3A and 3B on the IRAC band 2 image shown in Figure 3.
 
Below we provide some specifics regarding flow properties.
The total mechanical luminosity of the L1448-mm flow  is $\sim 1.5-3.5$ L$_{\odot}$ \citep{bac95, nisini00}.  The mass loss rate  has some uncertainty.  
\citet{nisini00} estimate $\dot{M}_{w} \sim 1 \times 10^{-6}$ M$_{\odot}$ yr$^{-1}$ from far-IR cooling lines.   Recent {\it Spitzer} data imply $\sim 10^{-6}$ M$_{\odot}$ yr$^{-1}$ and 10$^{-7}$ M$_{\odot}$ yr$^{-1}$ from the atomic and molecular flows, respectively \citep{dio09}.  The total H$_2$ v = 1 -- 0 S(1) luminosity from the L1448-mm flow is $\sim 0.19$ L$_{\odot}$ \citep{bll93}.   A comparison of the observed high-J CO, H$_2$O, and H$_2$ emission detected by ISO suggests that the flow is consistent with arising in low velocity C-shocks \citep[v $<$ 20 km s$^{-1}$; ][]{nisini00}, but at the tip of the bow shock higher velocities ($\sim 100$ km s$^{-1}$) are required \citep{dio09}.
\citet{nisini00} estimate that A$_V$ $\sim 6-9$  mag towards IRS3, A$_V \sim 4-6$ mag towards the tip of the L1448-mm flow,  and $\sim 4.7-7$ mag in the southern lobe.  The  distance of the source is $\sim$ 250 pc \citep{enoch06}.

\subsection{VLA1623}

The VLA1623 outflow was discovered by CO observations of cm-wave point sources coincident
with $\rho$ Oph core A \citep{ma88, andre90, lww90}.   The molecular data  revealed a remarkably
collimated outflow, which was notable for the clear lack of infrared point sources
in the region.   The detection of cold, centrally-concentrated sub-mm continuum emission led this source to be 
known as the prototypical Class 0 source \citep{andre93}.
Studies of the outflow itself find $\sim$0.1 M$_{\odot}$ entrained in the CO outflow \citep{dent95} with an estimated wind mass loss rate of $\dot{M}_{w} \sim 10^{-6}$ M$_{\odot}$ yr$^{-1}$ and mechanical luminosity of $\sim 1.9$ L$_{\odot}$\citep{bontemps96, andre90}.
The CO flow is coincident with a series of knots seen in images of H$_2$ vibrational emission,
with a total luminosity of L$_{\rm H_2}$[1-0 S(1)] $\sim$ 0.04 L$_{\odot}$ \citep{dent95,cog06}.

In Figure 5 we show a finder chart showing the VLA1623 outflow as delineated by  these known knots of H$_2$ emission.  The ``H'' sources follow the nomenclature of \citet{dent95}, but we also list the A (H5) and B (H4) sources as denoted by \citet{de95}.  Additional sources in the field have been found by a deeper H$_2$ survey by \citet{gomez03}.
We note that source H5 is also known as HH313 and that nearby in the field is the reflection nebula associated with the Class I source GSS30 \citep{gss73}.  The rectangles show the area mapped in our Spitzer IRS observations, which cover
the northwest red-shifted lobe of the flow encompassing from VLA1623 extending to  knots H4 and H5.

\citet{davis99} obtained high spectral resolution observations of H$_2$ S and Q vibrational lines with a focus on knot A (HH313, H4).   The resolved lines allowed an exploration of the ortho/para ratio of the hot shocked H$_2$ gas and they derived an o/p ratio of 3:1, consistent with thermal equilibrium at an estimated gas temperature in excess of 2000 K \citep[see also ][]{esd00}.
In addition, through the use of line ratios between lines with similar upper state energies, they estimate a K-band extinction of $A_k = 3 \pm 1$.
Recent VLBA parallax measurements of young stars associated with $\rho$ Oph A find a distance
of 120 pc \citep{loinard08}.

\section{Observations and results}

The five outflow sources discussed in \S 2 above were observed with the Short-Low (SL), Short-High (SH) and Long-High (LH) modules of the IRS instrument, providing coverage of the entire 5.2 -- 37 $\mu$m spectral region accessible to IRS at the maximum spectral resolving power available: $\lambda/\Delta \lambda \sim 60 - 127$ for SL and $\lambda/\Delta \lambda \sim 600$ for SH and LH.  Regions of size $\sim 1^\prime \times 1^\prime$ were mapped by stepping the IRS slit perpendicular to its long axis
by one-half its width and, in the case of SH and LH, parallel to its
long axis by 4/5 (SL) or 1/5 (LH) of its length. In each object, several such $\sim 1^\prime \times 1^\prime$ regions were arranged along the outflow axis to provide optimal coverage of the interaction of the outflow and the surrounding gas.  In the case of the SH and LH modules, we obtained spectra of offset regions that are devoid of strong molecular emission; these were used for background subtraction.  Table 1 summarizes the observational
details for each object.   With the exception of a single $\sim 1^\prime \times 1^\prime$ region mapped previously towards each of BHR71 and NGC 2071, the observations were carried out in Cycle 4 of the {\it Spitzer} mission

The data were processed at the Spitzer Science Center (SSC),
using version 17.2 or 18.0 of the processing pipeline.  They were then reduced
further using the SMART software package (Higdon et al. 2004),
supplemented by additional routines (Neufeld et al. 2006, 2007), which
provide for the removal of bad pixels in the LH and SH data, the
calibration of fluxes obtained for extended sources, the extraction
of individual spectra at each position sampled by the IRS slit, 
the creation of spectral-line maps from the set of extracted spectra, 
the removal of striping resulting from variations in response from one pixel 
to another, and -- in the case of the SH and LH modules -- the subtraction of a 
background spectrum obtained at an offset position devoid of strong molecular
emissions.

In Figures 6 - 10, we present, for each object, an example spectrum obtained by averaging the data over a simulated Gaussian beam of radius $25^{\prime\prime}$.  The beam center positions, specified in Table 1, were chosen to encompass regions of strong line emission but weak or moderate continuum emission.  The wavelength range plotted in Figure 6 - 10 is limited to 5.2 - 31 $\mu$m; the quality and reliability of the spectra diminish rapidly longward of 31~$\mu$m, particularly in objects that show a strong continuum flux. In the SL modules, each example spectrum shows a clear detection of the H$_2$ S(2) -- S(7) pure rotational lines; in the SH module, H$_2$ S(1) and S(2) are invariably observed with large line-to-continuum and signal-to-noise ratios; while in the LH module,
H$_2$ S(0) is typically detected, along with the $\rm ^3P_1 - ^3P_2$ 25.25$\rm\,\mu m$ fine structure transition of atomic sulphur and the $^6D_{7/2} - ^6D_{9/2}$ transition of singly-ionized iron.

Spectral line maps are shown in Figures 11 -- 15.  The effective angular resolution varies from $\sim 3^{\prime\prime}$ (HPBW) for the SL module to $\sim 10^{\prime\prime}$ for the LH module (see Table 5 of N06).  For each object, we present maps of the H$_2$ S(0),  
H$_2$ S(1), H$_2$ S(2), H$_2$ S(3), H$_2$ S(5), [FeII] 25.99 $\mu$m and [SI] 25.25 $\mu$m lines, together with the sum of the H$_2$ S(0) -- S(7) emission lines and the archival 3.6 $\mu$m, 4.5 $\mu$m, 5.8 $\mu$m, and 8 $\mu$m IRAC  maps.  The H$_2$ S(4), S(6) and S(7) line maps -- although somewhat noisier and not shown in Figures 11--15 -- all exhibit a morphology nearly identical to those of the H$_2$ S(3) and S(5) maps.  In the present paper, we consider only those spectral lines that can be mapped reliably over most of the region sampled by our observations.  Additional lines of H$_2$O, OH and HD -- typically much weaker than those considered here -- have been detected in parts of some our target objects (e.g. Melnick et al.\ 2008), and will be discussed further in upcoming papers. 

Each map is normalized relative to the reference intensities listed in Table 2.  For the spectral line maps, these reference intensities are not the maximum intensities but instead the values that are exceeded in only 0.1$\%$ of the mapped area\footnote{This choice of color scale allows the lower intensity emission to be discerned more readily.}; they are therefore somewhat smaller than the maximum measured intensities.  
For the continuum maps, the reference intensities were chosen to bring out the low intensity regions within the maps; they are often much smaller than the maximum intensities.  In the spectral line maps of BHR71, a $25^{\prime\prime}$ region near the map center has been excised from observations with the LH module;  here, the continuum intensity is too great to permit a reliable determination of the intensities of lines observed with the LH module.  For each object, the map center is listed in Table 1, and the tick marks indicate offsets relative to that position, small tick marks being separated by 10$^{\prime\prime}$ and large tick marks by 50$^{\prime\prime}$.  North is up and East is to the left.  The green circle plotted on the H$_2$ S(5) map indicates the simulated Gaussian beam for which the example spectra in Figures 6 -- 10 were obtained.

By integrating the line intensities over the entire mapped region for each object -- but excluding the excised region in BHR71 for those lines detected with the LH module -- we obtained the total line luminosities listed in Table 3.  These assume the distance estimates discussed in \S2 and tabulated in Table 1, and assume the line emission to be isotropic, as expected for optically thin lines.

\section{Discussion}

\subsection{H$_2$ rotational emissions}

The pure rotational transitions of H$_2$ are optically thin, and thus the level populations are readily calculable.  Our observations of the S(0) to S(7) rotational emissions provide estimates of the $J=2$ through $J=9$ column densities.  In this paper, we consider only the total H$_2$ luminosities for the entire mapped region in each object.  We defer a detailed discussion of spatial variations in the line luminosities and line ratios to a future paper, (Nisini et al.\, 
\re{ in preparation), in which we will also discuss existing spectral line maps of H$_2$ vibrational emissions.}  In Figure 16, we present rotational diagrams based upon the H$_2$ luminosities given in Table 3, corrected for extinction using the extinction estimates given in Table 1.  Here, we plot the logarithm of the total number of molecules in each state, divided by the degeneracy of that state, as a function of the energy.  Diamonds indicate the observed values, and the solid lines are results obtained from the fitting procedure described below.   While material in thermal equilibrium at a single temperature would exhibit a linear behavior in such a plot, the observed rotational diagrams -- like those obtained for outflow sources by N06 and by Maret et al.\ (2009), and for supernova remnants by N07 -- depart from that behavior in two respects.  First, there is a positive curvature to the rotational diagram, indicative of the presence of multiple temperature components.  Second, except toward L1448, there is a characteristic zigzag behavior, most notably for the states of lowest energy, in which the states of even rotational quantum number $J$ lie systematically above those of odd $J$.  As in N06, N07, and Maret al.\ (2009), we attribute this second behavior to an ortho-to-para ratio -- smaller than the equilibrium value of 3 -- that is the relic of an earlier period when the gas was considerably cooler than it is now.

We have modeled the H$_2$ rotational diagrams in Figure 16, using a method similar to that described by Neufeld and Yuan (2008; hereafter NY08).   As in NY08 
\re{ (their eqn.\ 3)}, we assume a power law distribution of gas temperatures with the number of molecules at temperature between $T$ and $T+dT$ assumed proportional to $T^{-b}$.  We assume a temperature range extending from $T_{min} = 100$~K \footnote{Recognizing that gas in the 100 -- 300~K can contribute significantly to the pure rotational line emission that we detect with {\it Spitzer}, we have reduced $T_{min}$ from 300~K in NY08 to 100~K in the present work.} to $T_{max}=4000$~K.  Solving the equations of statistical equilibrium for the H$_2$ level populations, and adopting the molecular data cited in NY08, we obtained predictions for the pure rotational line luminosities as a function of the power law index, $b$, the total mass of hydrogen molecules warmer than $T_{min}$, $M_{\rm H2}(\ge T_{min})$, and the H$_2$ density, $n({\rm H}_2)$.  
\re{ The excitation of pure rotational emissions is dominated by collisions with H$_2$ (see NY08), in contrast to the case of rovibrational emissions, where excitation by atomic hydrogen presents an additional complication that we will discuss in a future paper (Nisini et al.\ 2009, in preparation)}. 

We have extended the treatment of NY08 by considering the H$_2$ ortho-to-para ratio along with $b$, $M_{\rm H2}(\ge T_{min})$ and $n({\rm H}_2)$.  Here, we introduce two additional parameters: the initial ortho-to-para ratio, OPR$_0$, and the time period, $\tau$, for which the gas temperature has been elevated.  If reactive collisions with atomic hydrogen are the dominant mechanism for the interconversion of ortho- and para-H$_2$, then the ortho-to-para ratio after time $\tau$ is given by N06\footnote{Note that the expression given by N06 erroneously omits $k_{\rm OP}$ from the term $[k_{\rm PO}+k_{\rm OP}]$, resulting in a 25$\%$ error.}:
$${\rm OPR(\tau) \over 1+OPR(\tau)} =  {\rm OPR_0 \over 1+OPR_0} \, e^{-n({\rm H})\, \tau \, [k_{\rm PO}+k_{\rm OP}]} +  {\rm OPR_{LTE} \over 1+OPR_{LTE}} \, \biggl( 1- e^{-n(H)\, \tau \, [k_{\rm PO}+k_{\rm OP}]}\, \biggr), \eqno(1)$$
where $n({\rm H})$ is the atomic hydrogen density, $k_{\rm PO}$ is the rate coefficient for para-to-ortho conversion, estimated as $8 \times 10^{-11} \exp(-3900/T) \rm \, cm^3 \, s^{-1}$ by Schofield et al.\ (1967), based upon the laboratory experiments of (Schulz \& Le Roy 1965), and $k_{\rm OP} \sim k_{\rm PO}/3$ is the rate coefficient for ortho-to-para conversion.  Thanks to the strong dependence of $k_{PO}$ upon the gas temperature, the ortho-to-para ratio is similarly temperature-dependent, the warmer gas components being closer to equilbrium than the cooler ones. 
This effect is evident in Figure 16, the degree of zigzag being least for the highest energy states that are excited primarily in the warmest gas.

We have adjusted the five parameters $b$, $M({\rm H}_2)$, $n({\rm H}_2)$, OPR$_0$, and $n({\rm H})\tau$ in order to obtain the best fit to the data.  Here, we minimize the $\chi^2$ for the S(1) through S(7) rotational lines, assuming the same fractional uncertainty for each data point.  We exclude the H$_2$ S(0) line flux from the analysis, because its value is rather uncertain, particularly in regions of strong continuum intensity where the line-to-continuum ratio becomes small.
The best fit parameters are given in Table 3.  

The inferred masses of warm\footnote{
\re{ In this paper, we use the word ``warm" to describe H$_2$ at temperatures $\ge 100$~K.  This, of course, includes gas at lower temperatures than that which can be probed by means of traditional, ground-based studies of vibrationally-excited H$_2$.}}  H$_2$ (i.e.\ at $T \ge 100$~K) range from $1.3 \times 10^{-3} M_\odot$ in VLA 1623 to $0.18 M_\odot$ in NGC 2071.  These values exclude the mass contribution from helium and apply only to the regions mapped by IRS.  The best-fit power-law indices, $b$, range from 2.2 to 3.3.   These values lie somewhat below the typical values determined by NY08 in the SNR IC443C, and somewhat below the value $b = 3.8$ expected for a paraboloidal bow shock that is dissociative at its apex.  The best fit H$_2$ densities range from $3000$ to $7000\,\rm cm^{-3}$.  It should be noted, however, that the range of acceptable values for $b$ and $n({\rm H}_2)$ can be large, and that these two parameters are somewhat degenerate in the following sense: increasing $b$ tends to decrease the predicted relative strength for the higher $J$ transitions, whilst decreasing $n(\rm H_2)$ has a similar effect (because the level populations become increasingly subthermal).  Thus, an increase in one of these assumed parameters can partially compensate for a decrease in the other.  The correlated uncertainties in $b$ and $n(\rm H_2)$ are presented graphically in Figure 17, which shows contours of $\chi^2$ in the $b$--$n(\rm H_2)$ plane.  Crosses indicate the minimum $\chi^2$ (i.e. the best fit), while the two contours represent the 68$\%$ and 95$\%$ confidence limits on each parameter.  With the exception of L1448, for which the data are consistent with an OPR of 3, the objects require an initial OPR less than 3.  In two objects, L1157 and NGC 2071, the current OPR is clearly larger for the higher-$J$ states than for the lower-$J$ states.  In the context of our model for the evolution of the OPR (eqn.\ 1), both cases imply a lower limit $\sim 10^3 \,[\rm{cm}^{-3}/n({\rm H})]$~yr on the time period, $\tau$, for which the gas has been warm.

\subsection{Fine structure emissions}

The Fe$^+$ $\,\,\rm ^6D_{7/2} - ^6D_{9/2}  \,\,\,\,25.988 \,\mu$m and S$\,\,\,\,\rm ^3P_{1} -^3P_{2} \,\,\,\, 25.249 \,\mu$m fine structure lines are readily detectable in the LH spectra of all five outflows sources, and are strong enough to map.  However, these transitions make a negligible contribution to the total 5.2 -- 37 $\mu$m line luminosity.  The [SI] line luminosity ranges from $\sim$ 1 to 2 $\%$ of the total H$_2$ S(0) -- S(7) luminosity.  For [FeII], the corresponding range is wider ~ $\sim$ 0.15 to 2$\%$.  In the case of [FeII], these percentages are much smaller than those attained in some of the supernova remnants (SNR) that we observed previously (N07).  In the SNR 3C391, for example, the [FeII] $25.988 \,\mu$m line flux exceeded 10$\%$ of the total H$_2$ S(0) -- S(7) flux.  Furthermore, fine structure emissions from more highly ionized species -- Ne$^+$, Ne$^{++}$, and S$^{++}$ for example -- are readily detectable in SNR but not in protostellar outflow sources.  These differences suggest that higher velocity shocks are more prevalent in supernova remnants than around outflow sources.

A visual inspection of Figures 11 -- 15 suggests that the [SI] emissions are typically more closely correlated with the H$_2$ emissions than are the [FeII] emissions.  This impression is borne out by a correlation analysis.  Comparing the H$_2$~S(5) and [SI] intensities, spatial pixel by spatial pixel, we obtained linear correlation coefficients of 0.81, 0.78, 0.42, 0.80, and 0.85 respectively for BHR71, L1157, L1448, NGC 2071 and VLA 1623.  Comparing the H$_2$~S(5) and [FeII] intensities, the corresponding results are invariably smaller: 0.55, 0.48, 0.23, 0.72, and 0.78.  This behavior is consistent with our previous study of supernova remnants, in which a principal component analysis suggested that the [SI] emitting region was largely cospatial with the H$_2$ emitting region, while emissions from ionized species such as Fe$^+$ originated in a separate gas component.  This segregation is presumably a consequence of the different shock velocities needed to generate [SI], H$_2$, and [FeII] emissions.  The first two arise in slow non-dissociative shocks that create very little ionization, while the third arises in faster shocks that are dissociative and ionizing.  According to theoretical models (e.g.\ Allen et al.\ 2008), the emission from dissociative shocks is dominated by optical and ultraviolet line emissions, with infrared fine structure emissions accounting for only a small fraction of the overall shock luminosity.

\subsection{Source energetics and comparison with IRAC maps}

The wavelength coverage of {\it Spitzer} is well suited to observing the radiative cooling of non-dissociative molecular shocks.  The study of Kaufman and Neufeld (1996) indicates that over a wide range of shock parameters, pure rotational emissions from H$_2$ are key coolants of non-dissociative shocks.  For shocks of velocity in the range 20 to 40~km~s$^{-1}$ and with a preshock H$_2$ density of $10^4$~cm$^{-3}$, for example, between 45 and 69$\%$ of the shock luminosity is predicted to emerge in pure rotational emission from H$_2$.  Moreover, almost all of the H$_2$ pure rotational emission emerges in the S(0) - S(7) lines accessible to {\it Spitzer}.    The total H$_2$ S(0) -- S(7) luminosities listed in Table 1 range from 0.02 to 0.75 $\, L_\odot$.   Taking these luminosities as lower limits for the total mechanical luminosity dissipated in the shocks, adopting 20~km~s$^{-1}$ as a typical shock velocity, $v_s$, and setting the mechanical luminosity equal to  $\onehalf \dot{M} v_s^2$, we obtain lower limits on the mass flow rates through the shock, $\dot{M}$,  ranging from $\sim 10^{-6}$ to $\sim 2 \times 10^{-5} M_\odot \rm \, yr^{-1}$.  

In some objects, H$_2$ line emissions can contribute significantly to the intensities measured using broadband mid-IR photometry (e.g. Reach et al.\ 2006, NY08).  For the supernova remnant IC443, NY08 found that the mid-IR spectral region covered by IRAC was dominated by H$_2$ line emissions.  In particular, the intensities measured in the 5.8~$\mu$m and 8~$\mu$m IRAC channels (Bands 3 and 4), were accounted for almost entirely by the IRS-measured  H$_2$ S(4) - S(7) line intensities.  Given an excitation model for H$_2$, the 3.6 $\mu$m and 4.5 $\mu$m IRAC intensities could similarly be explained as resulting from H$_2$ emissions, primarily the S(9) pure rotational emission for the 4.5 $\mu$m band and the v = 1 -- 0 Q(5) rovibrational line for the 3.6 $\mu$m band.  

For the protostellar outflow sources studied here, we can evaluate the contribution of the H$_2$ S(4) -- S(7) emissions to the broadband intensities measured with IRAC.  In Table 3, we list the fraction of the intensities measured in the 5.8~$\mu$m and 8~$\mu$m IRAC bands that is attributable to H$_2$ S(4) and S(5) 
(in the case of the 8~$\mu$m band) or H$_2$ S(6) and S(7) (in the case of the 5.8~$\mu$m band).  
The contribution is clearly largest in BHR71 and L1157, for which the computed contribution to the 8~$\mu$m band actually exceeds 100$\%$, a discrepancy that presumably reflects flux calibration errors that are within expectations.  Even in these objects, however, the 5.8~$\mu$m band intensity is 2--3 times stronger than what can be attributed to H$_2$, and in the three other objects, the contribution of H$_2$ to either the 5.8~$\mu$m or 8.0~$\mu$m band is small, at least averaged over the entire mapped region.  Nevertheless, the IRAC maps shown in Figures 11 - 15 can exhibit clear morphological similarities to the H$_2$ line maps, even in objects such as L1448 where the overall contribution of H$_2$ is small.  This reflects the fact that regions of diffuse emission can be H$_2$-dominated even if compact continuum sources account for most of the IRAC intensity when averaged over the entire region that we have mapped.  Finally, we note that in the case of VLA 1623 and NGC 2071, the contributions listed in Table 3 reflect an important selection effect: we specifically avoided observing H$_2$ emissions in those regions known to exhibit the strongest continuum emission, because they would have saturated the IRS detectors.  Thus, had we been able to map the entire object with IRS, the fractional contributions of H$_2$ listed in Table 3 would have been even smaller, probably by a large factor.  In summary, a comparison of the IRS and IRAC data indicates that the 8$\mu$m IRAC channel can be dominated by H$_2$ emissions for a protostellar outflow, as it was for the supernova remnant IC443 discussed by NY08, but that the H$_2$ contribution can also be negligible, as in NGC 2071 and VLA 1623.

\section{Summary and future outlook}

1) We have carried out spectroscopic mapping observations toward protostellar outflows in the BHR71, L1157, L1448, NGC 2071, and VLA 1623 molecular regions using the Infrared Spectrograph (IRS) of the {\it Spitzer Space Telescope}.  

2) Our observations provide detailed maps of the 8 lowest pure rotational lines of molecular hydrogen and of the [SI] 25.25$\rm\,\mu m$  and [FeII] 26.0$\rm\,\mu m$ fine structure lines. 

3) Within the regions mapped towards these 5 outflow sources, total H$_2$ luminosities ranging from 0.02 to 0.75 L$_{\odot}$ were inferred for the sum of the 8 lowest pure rotational transitions.  The overall H$_2$ luminosities were fit with a model in which an admixture of gas temperatures is present within the beam.  For an assumed power-law distribution of gas temperatures, with the mass of material at temperature $T$ to $T+dT$ assumed proportional to $T^{-b} dT$, the best-fit power-law indices lie in the range 2.3 to 3.3.  With the exception of L1448, all objects exhibit a non-equilibrium H$_2$ ortho-to-para ratio $\le 3$.  L1157 and NGC 2071 show evidence for an H$_2$ ortho-to-para ratio that is an increasing function of gas temperature, as expected if para-to-ortho conversion possesses an activation energy barrier. 

4) As in supernova remnants studied by Neufeld et al.\ 2007, the [SI] 25.3$\,\mu$m fine structure emission is more strongly correlated with the H$_2$ emissions than is the [FeII] 26.0$\rm\,\mu m$ fine structure emission.  This behavior is consistent with the suggestion that the [FeII] 26.0$\rm\,\mu m$ fine structure transition traces faster, dissociative shocks.

5) In BHR71 and L1157, H$_2$ S(4) and S(5) line emissions account for most of the intensity observed in IRAC 8$\mu$m maps of the region we mapped with IRS.

The primary emphasis in the present paper has been on object-averaged line luminosities.  However, the data presented here will allow the construction of line ratio maps that can probe the {\it variation} of the physical conditions within each individual object; these will be discussed in a series of future papers.

\begin{deluxetable}{lccccc}
\rotate
\tablewidth{0pt}
\tablecaption{Observational details and source properties} 
\tablehead{& BHR71 & L1157 & L1448 & NGC 2071 & VLA 1623}
\startdata

Adopted distance (pc) 	& 200 & 440 & 250 & 390 & 120 \\
Map center R.A. (J2000) & 12h 01m 36.07s & 20h 39m 5.11s & 3h 25m 38.45s & 5h 47m 4.05s & 16h 26m 22.96s\\
Map center Dec. (J2000) & 
--65d 08$^{\prime}\,50.5^{\prime\prime}$&
68d 02$^{\prime}\,43.9^{\prime\prime}$&
30d 44$^{\prime}\,13.8^{\prime\prime}$&
0d 21$^{\prime}\,56.3^{\prime\prime}$&
--24d 23$^{\prime}\,56.2^{\prime\prime}$ \\
Reference position R.A. $^a$ & 12h 03m 40.00s & 20h 38m 03.00s & 3h 24m 40.00s & 5h 46m 50.00s & 16h 26m 11.00s\\
Reference position Dec. $^a$ & 
--64d 58$^{\prime}\,20.0^{\prime\prime}$&
68d 04$^{\prime}\,0.0^{\prime\prime}$&
30d 46$^{\prime}\,0.0^{\prime\prime}$&
0d 11$^{\prime}\,55.0^{\prime\prime}$&
--24d 17$^{\prime}\,50.0^{\prime\prime}$ \\

Start date for IRS observations & 2005 Jul 02 & 2007 Oct 31 & 2008 Feb 19 & 2004 Oct 21 & 2008 Mar 27\\
End date for IRS observations   & 2008 Feb 29 & 2007 Nov 01 & 2008 Feb 29 & 2008 Mar 27 & 2008 Mar 28\\
IRS observing time (hr)		& 18.8	      & 22.8        & 22.8        & 23.9            & 18.3       \\
Beam offset$^b$ for plotted spectrum & (--2, 78) & (+28, --85) & (34, --82) & (63, 59) & (--53, 51)\\
Extinction estimate (A$_V$ in mag)   & 2$^c$	 &	2$^c$     &     6$^d$      &    13$^e$    & 15$^{c,f}$\\ 
\enddata
\tablenotetext{a}{Used for background subtraction for the SH and LH modules}
\tablenotetext{b}{($\Delta \alpha \cos \delta, \Delta \delta$) relative to the map center in units of arcsec}
\tablenotetext{c}{Caratti o Garatti et al.\ 2006}
\tablenotetext{d}{Nisini et al.\ 2000}
\tablenotetext{e}{Melnick et al.\ 2008}
\tablenotetext{f}{Davis et al.\ 1999}
\end{deluxetable}
\clearpage
  
\begin{deluxetable}{lrrrrr}
\tablewidth{0pt}
\tablecaption{Reference intensities for each map$^a$} 
\tablehead{& BHR71 & L1157 & L1448 & NGC 2071 & VLA 1623}
\startdata

H$_2$ S(5) \,\,  6.910$\,\mu$m     				& 372.1 & 262.0 & 256.8 &1290.0 & 409.3\\
H$_2$ S(3) \,\,  9.665$\,\mu$m     				& 239.6 & 256.3 & 225.7 & 628.3 & 183.5\\
H$_2$ S(2) \,\,  12.279$\,\mu$m     				& 53.4  &  96.7 &  60.8 & 300.3 & 133.8\\
H$_2$ S(1) \,\,  17.035$\,\mu$m     				& 34.5  &  55.0 &  49.4 & 164.2 &  34.1\\
H$_2$ S(0) \,\,  28.219$\,\mu$m     				& 4.1   &   8.0 &   6.0 &  31.8 &  15.7\\
H$_2$ S(0) - S(7) total            			      	&1225.0 & 984.0 & 921.8 &3912.0 &1377.0\\
Fe$^+$ $\,\,\rm ^6D_{7/2} - ^6D_{9/2}  \,\,\,\,25.988 \,\mu$m  	& 4.5   &   6.1 &  20.5 &  18.7 &  27.0\\
S$\,\,\,\,\rm ^3P_{1} -^3P_{2} \,\,\,\, 25.249   \,\mu$m 	& 25.0  &  12.3 &  17.0 &  59.5 &  27.3\\
IRAC Band 1 (3.6 $\,\mu$m cont.) & 10 &  3  &  5 &  40 & 100\\
IRAC Band 2 (4.5 $\,\mu$m cont.) & 20 &  3  & 10 & 100 & 100\\
IRAC Band 3 (5.8 $\,\mu$m cont.) & 40 & 10  & 10 & 200 & 300\\
IRAC Band 4 (8.0 $\,\mu$m cont.) & 20 &  5  & 10 & 200 & 500\\

\enddata

\tablenotetext{a}{in units of $10^{-6} \, \rm erg \, cm^{-2} \,s^{-1} \, sr^{-1}$ for lines and MJy$\rm \, sr^{-1}$ for continuum.  
\re{ The errors in the line intensities are likely dominated by systematic effects and are expected to be $\le 25 \%$ (N06).}}
\end{deluxetable}

\clearpage

\begin{deluxetable}{lrrrrr}
\tablewidth{0pt}
\tablecaption{Measured line luminosities and derived parameters} 
\tablehead{& BHR71 & L1157 & L1448 & NGC 2071 & VLA 1623}
\startdata
\underline{Line luminosities$^a$} \\
H$_2$ S(7) \,\,  5.511$\,\mu$m 					&  7.1 &  15.3  &  5.0 & 94.5 & 2.1 \\
H$_2$ S(6) \,\,  6.100$\,\mu$m     				&  3.9 &   8.5  &  3.4 & 54.4 & 2.1 \\
H$_2$ S(5) \,\,  6.910$\,\mu$m     				& 11.4 &  27.9  &  9.2 &206.4 & 5.6 \\
H$_2$ S(4) \,\,  8.025$\,\mu$m     				&  6.8 &  25.9  &  7.9 & 98.7 & 2.5 \\
H$_2$ S(3) \,\,  9.665$\,\mu$m     				& 10.1 &  42.1  & 10.8 &145.7 & 2.7 \\
H$_2$ S(2)$^b$ \,\,  12.279$\,\mu$m     				&  2.2 &  16.2  &  2.9 & 79.0 & 1.6 \\
H$_2$ S(1) \,\,  17.035$\,\mu$m     				&  1.7 &  13.0  &  3.7 & 58.2 & 0.6 \\
H$_2$ S(0) \,\,  28.219$\,\mu$m     				&  0.3 &   2.3  &  0.6 & 12.0 & 0.4 \\
H$_2$ S(0) - S(7) total            			      	& 43.7 & 151.2  & 43.6 &748.8 &17.7 \\
Fe$^+$ $\,\,\rm ^6D_{7/2} - ^6D_{9/2}  \,\,\,\,25.988 \,\mu$m  	&  0.5 &   0.2  &  0.7 &  2.9 & 0.3 \\
S$\,\,\,\,\rm ^3P_{1} -^3P_{2} \,\,\,\, 25.249   \,\mu$m 	&  0.7 &   1.4  &  0.8 &  8.8 & 0.3 \\
\\
\multicolumn{5}{l}{\underline{Contribution of H$_2$ to IRAC bands (percentage)}} \\
Band 3 (5.8 $\,\mu$m) &  33 &  39 & 15 & 9.1 & 6.5 \\
Band 4 (8.0 $\,\mu$m) & 120$^c$ & 106$^c$ & 18 & 5.7 & 2.6 \\
\\ 
\underline{H$_2$ excitation: best-fit parameters$^d$}\\
H$_2$ mass above 100 K ($10^{-3} M_\odot$)  	& 2.5  & 49 &  8.1 & 177  & 1.3\\
Power law index, b = -dln$M$/dln$T$ 		& $2.5\pm 0.5$  & $3.3 \pm 0.5$  &  $2.9 \pm 0.5$  & $3.2 \pm 0.6$  & $2.3_{-0.8}^{+1.3}$\\
log(H$_2$ density/cm$^{-3}$)	 		& $3.8\pm 0.4$  & $3.8 \pm 0.5$  &  $3.8 \pm 0.4$  & $3.8 \pm 0.4$  & $3.5_{-0.1}^{+0.7}$\\
Initial H$_2$ OPR (OPR$_0$) 			& $\le 2.4$  & $\le 1.8$  &  note e  & $\le 1.8$  & $\le 1.9$ \\

\enddata
\tablenotetext{a}{in units of $10^{-3} \, L_\odot$ for assumed distances of 200, 440, 250, 390, and 120~pc respectively for BHR71, L1157, L1448, NGC 2071, and VLA 1623.  These values do not include any extinction correction.  
The errors in the line intensities are likely dominated by systematic effects and are expected to be $\le 25 \%$ (N06).}

\tablenotetext{b}{The luminosity for H$_2$ S(2) is determined from the SL module}

\tablenotetext{c}{Here, the computed contribution to the 8~$\mu$m band actually exceeds 100$\%$, a discrepancy that presumably reflects flux calibration errors that are within expectations.}

\tablenotetext{d}{$\pm$ and $\le$ symbols indicate 95$\%$ confidence intervals}

\tablenotetext{e}{No useful constraint (consistent with 3 or any smaller value)}

\end{deluxetable}

\clearpage

\begin{figure}
\includegraphics[scale=0.80]{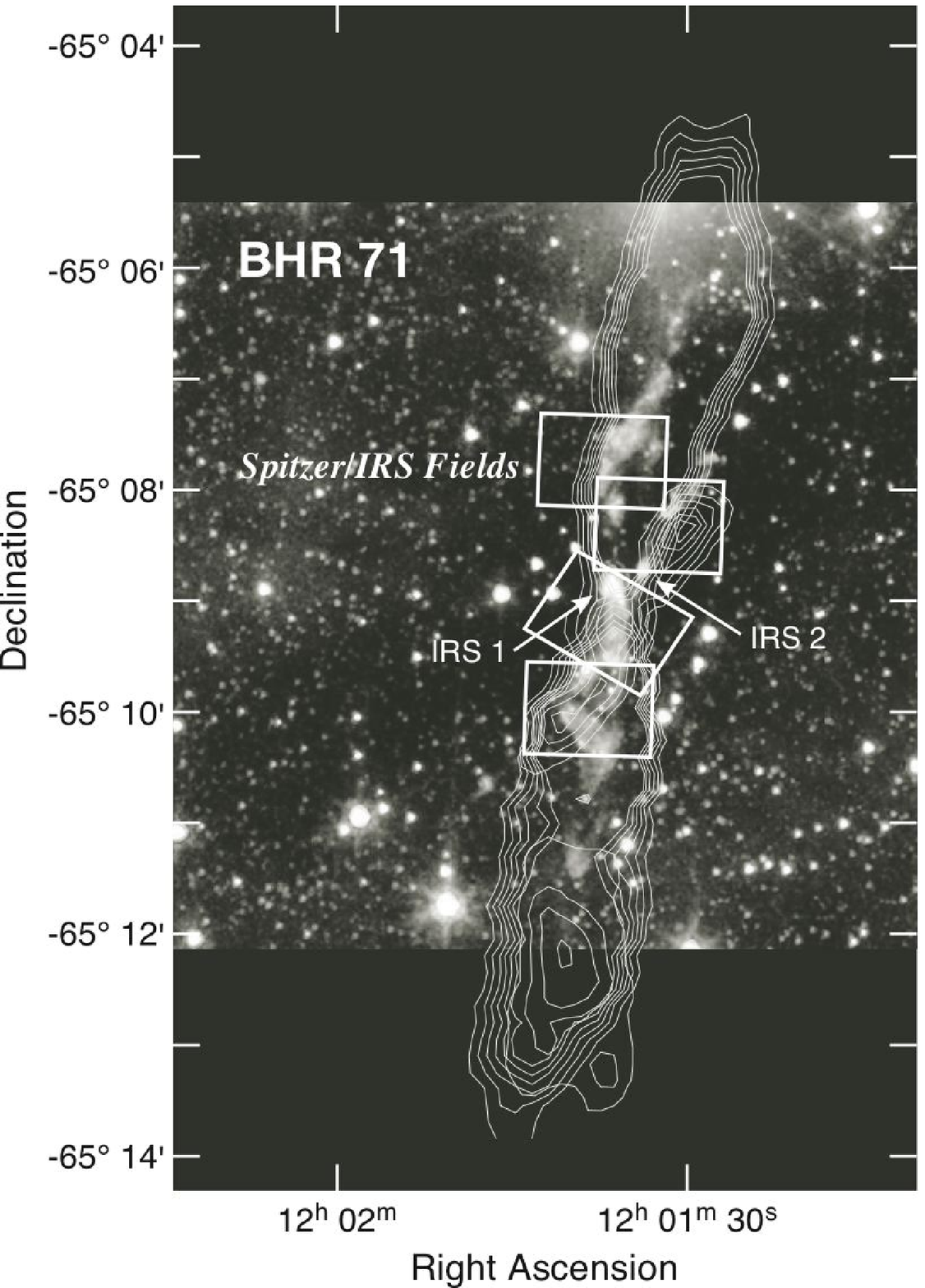}

\noindent{Fig.\ 1 -- IRAC four-band greyscale image of the BHR71 region
on which are superposed contours of integrated CO(3--2) emission
(Parise et al.\ 2006).  The positions of
the two Class 0 sources associated with BHR~71 are shown; IRS~1
is believed responsible for driving the north-south outflow, while
IRS~2 is driving a much smaller east-west outflow.  The rectangles 
denote the fields surveyed by {\em Spitzer}/IRS (for the LH module).}
\end{figure}

\begin{figure}
\includegraphics[scale=0.95]{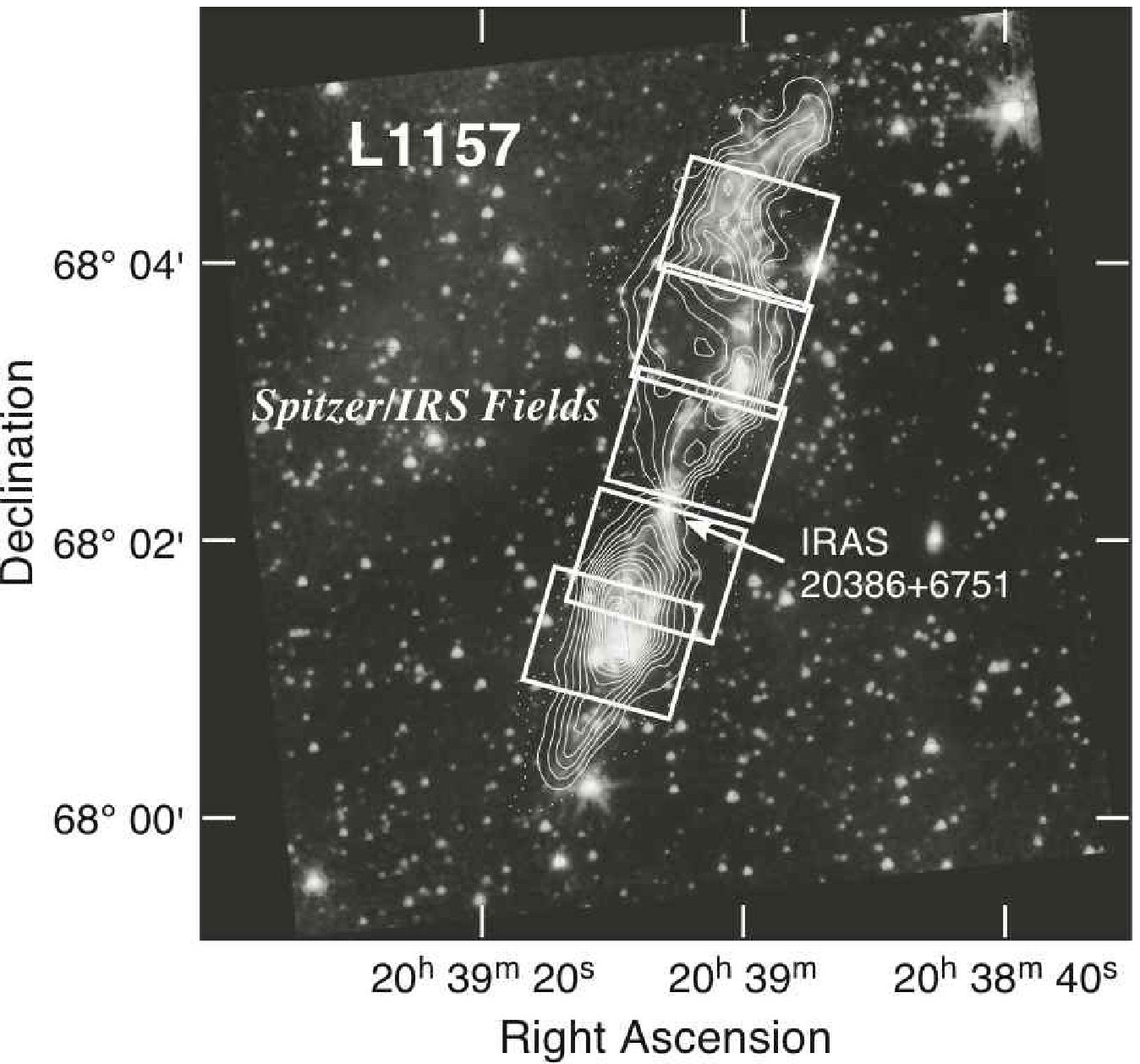}

\noindent{Fig.\ 2 -- IRAC greyscale image of the L1157 region.  
The color stretch is slightly exaggerated to emphasize
channel 4 (8$\,\mu$m), where the extinction is the largest.
Superposed on the {\em Spitzer} image is the CO (2--1) emission
from Bachiller et al. (2001).  The position of the driving source,
IRAS 20386+6751, is shown along with the fields surveyed by 
{\em Spitzer}/IRS (LH module).}
\end{figure}

\begin{figure}
\includegraphics[scale=0.95]{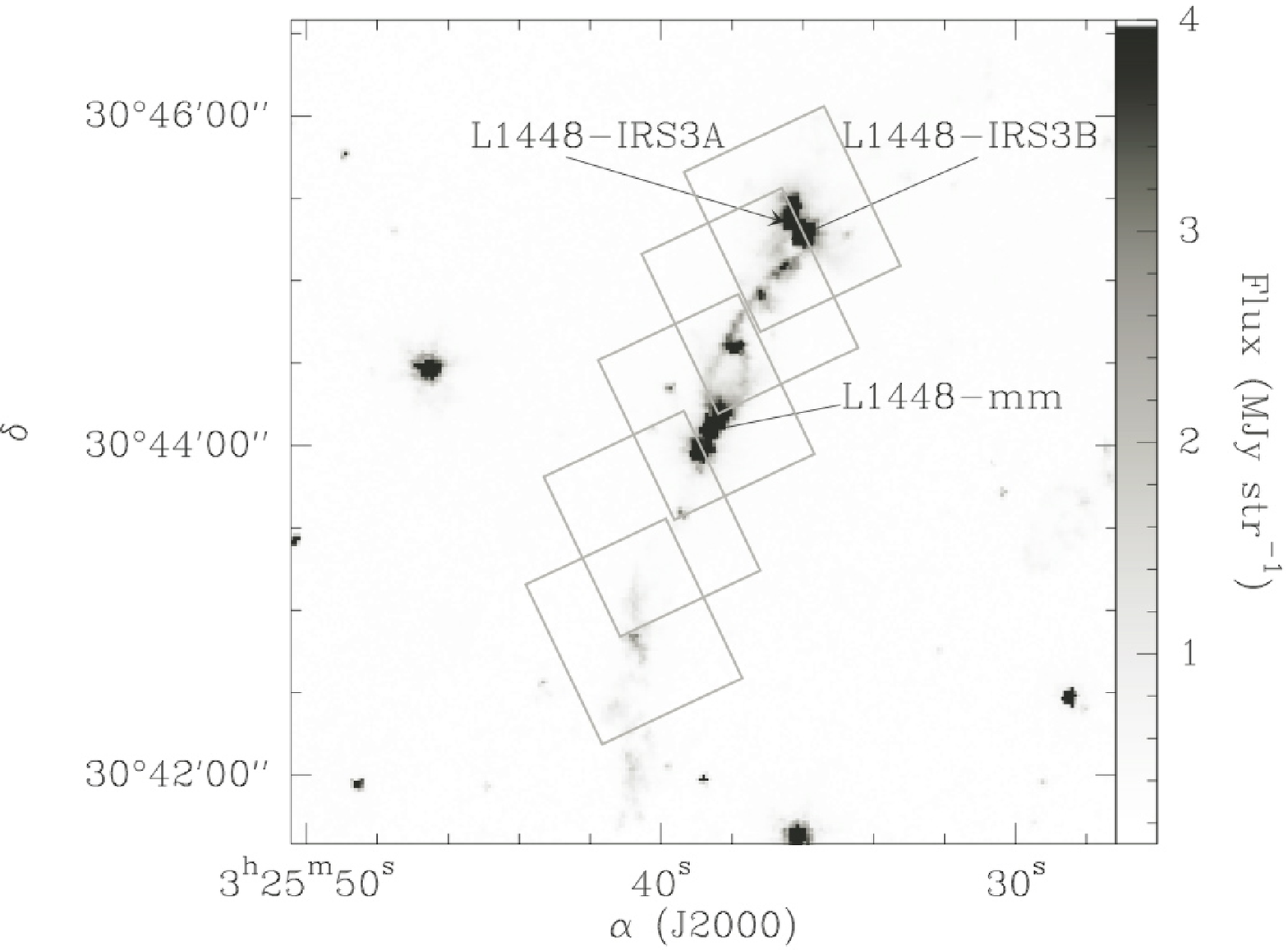}

\noindent{Fig.\ 3 -- Image of {\it Spitzer}/IRAC band 2 in the region surrounding the Class 0 source L1448.  Prominent sources are labelled in the image.
The rectangles denote the fields surveyed by {\em Spitzer}/IRS (LH module).}
\end{figure}

\begin{figure}
\includegraphics[scale=0.80]{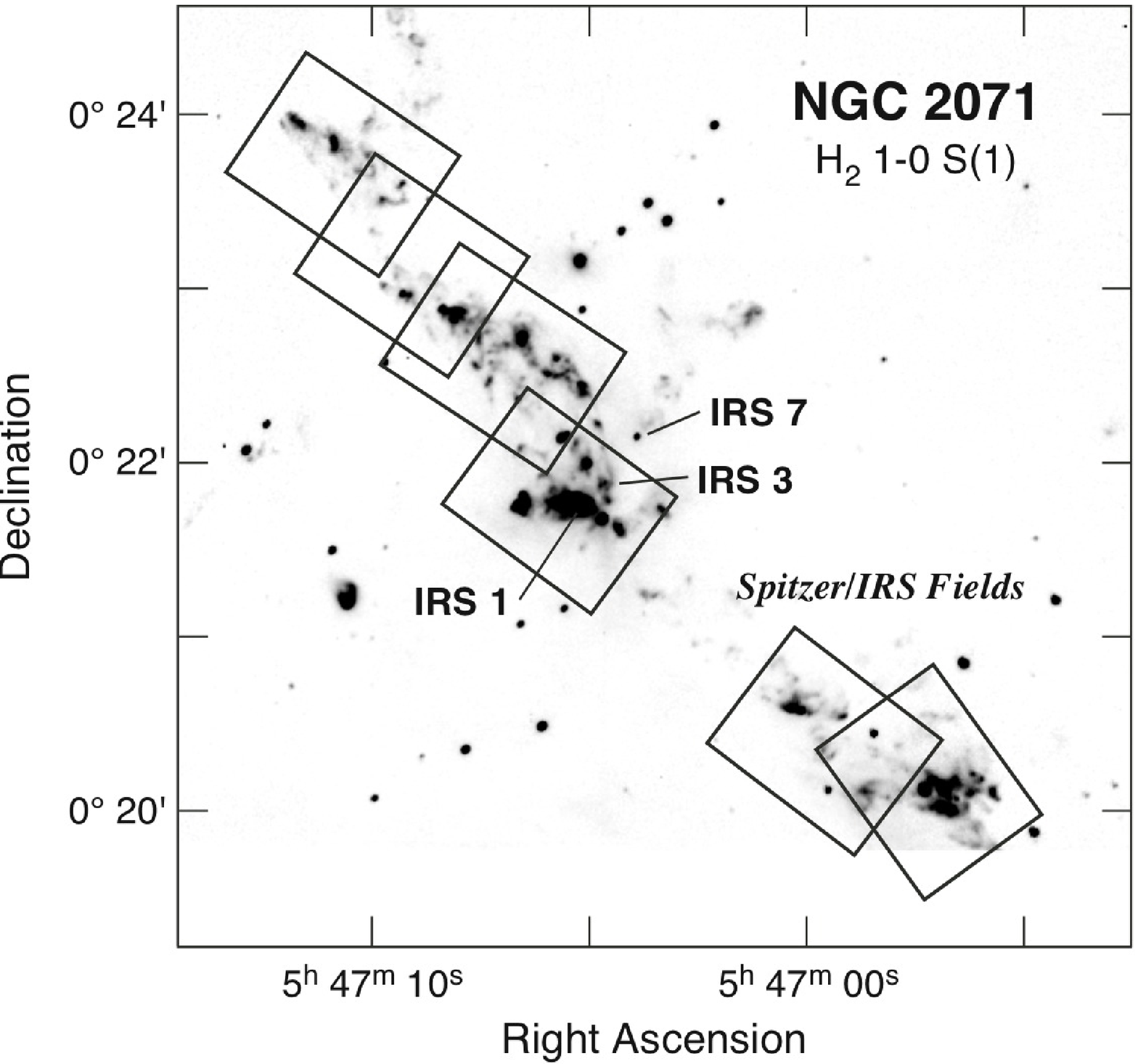}

\noindent{Fig.\ 4 --  Map of  the outflow from NGC 2071 as traced by the H$_2$ 1-0
S(1) 2.12 $\mu$m emission (Eisl\"offel 2000).  The rectangles 
denote the fields surveyed by {\em Spitzer}/IRS (LH module).}
\end{figure}

\begin{figure}
\includegraphics[scale=0.80]{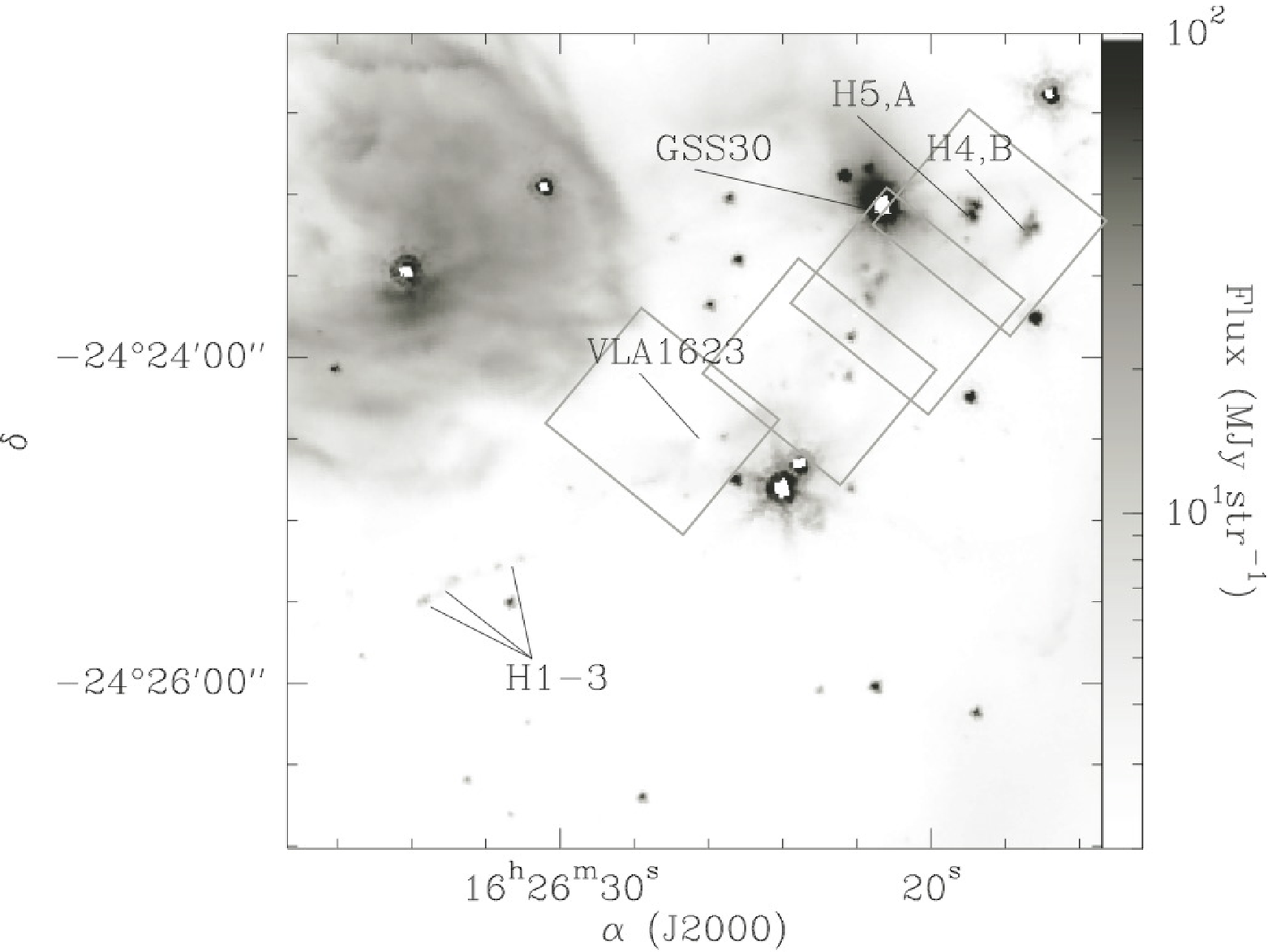}

\noindent{Fig.\ 5 -- Image of {\it Spitzer}/IRAC band 2 in the region surrounding the Class 0 source VLA1623.  Sources labeled with an H denote knots of vibrationally excited H$_2$.   
\re{ Source H5 is also known as HH313.}  The grey box shows the approximate coverage of our {\it Spitzer}/IRS maps.}
\end{figure}

\begin{figure}
\includegraphics[scale=1.0,angle=0]{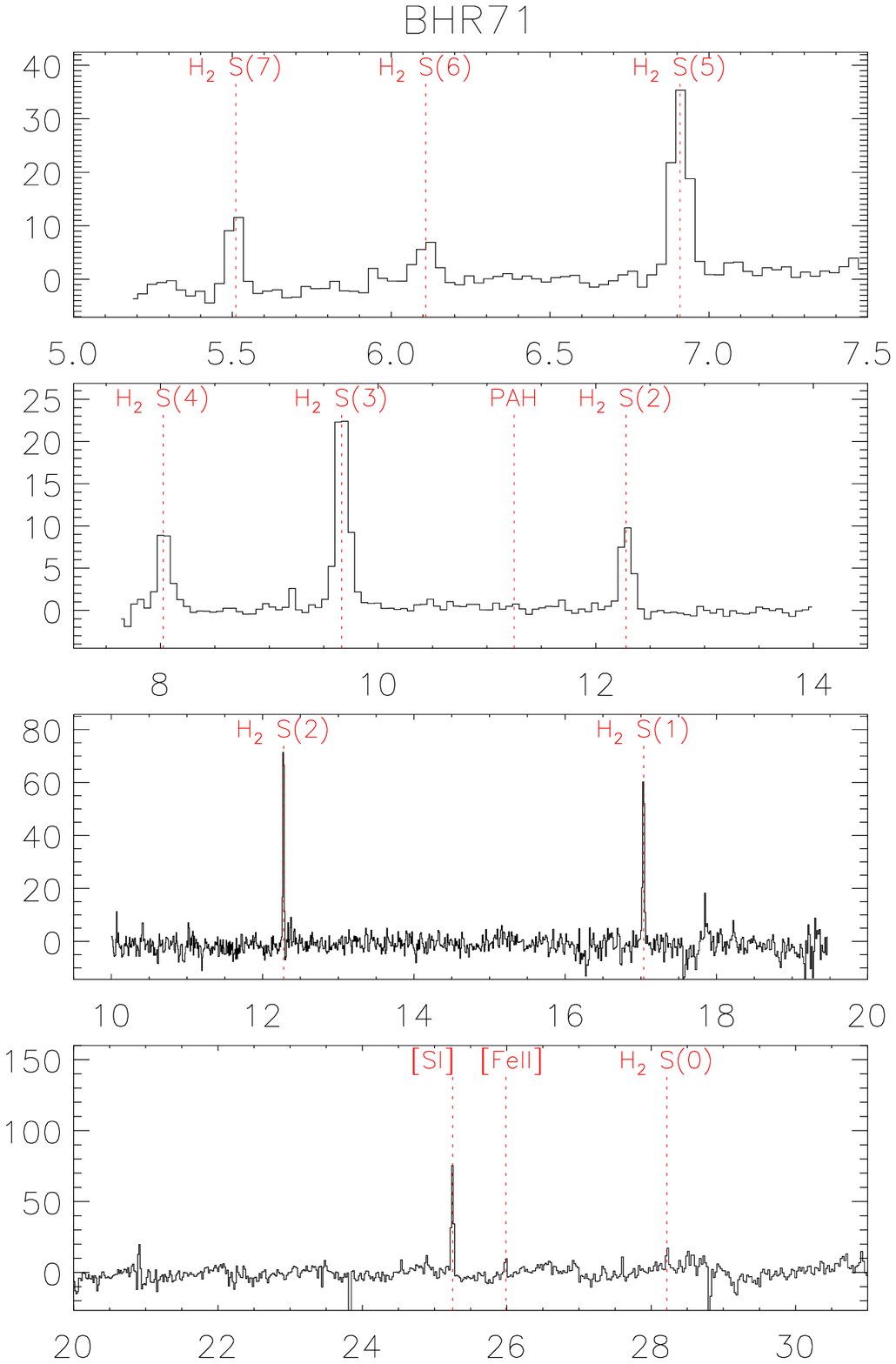}

\noindent{Fig.\ 6 -- Example spectrum for BHR71, obtained by averaging the data over a simulated Gaussian beam of radius $25^{\prime\prime}$.  The beam center position, 
\re{ $\alpha=$ 12h 01m 35.75s, $\delta=$ --65d 07$^{\prime}\,33.5^{\prime\prime}$ (J2000)}, was chosen to encompass a region of strong line emission but weak or moderate continuum emission.  The wavelengths are given in units of $\mu$m, and the intensities in units of MJy~sr$^{-1}$.}
\end{figure}

\begin{figure}
\includegraphics[scale=1.0,angle=0]{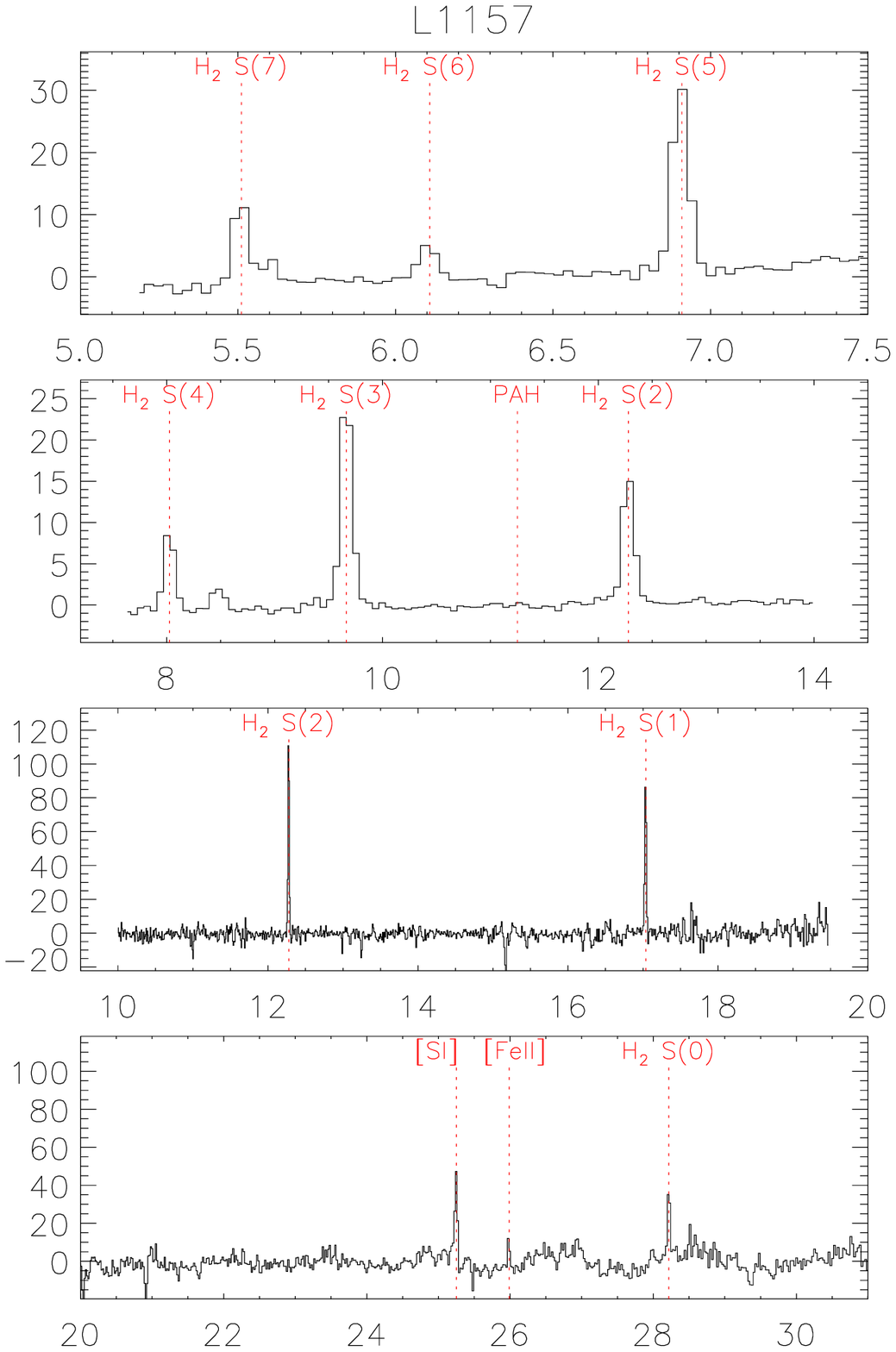}

\noindent{Fig.\ 7 -- same as Fig.\ 6, but for L1157.  
\re{ The beam center position is $\alpha=$ 20h 39m 10.09s, $\delta=$ 68d 01$^{\prime}\,18.9^{\prime\prime}$ (J2000) }}
\end{figure}

\begin{figure}
\includegraphics[scale=1.0,angle=0]{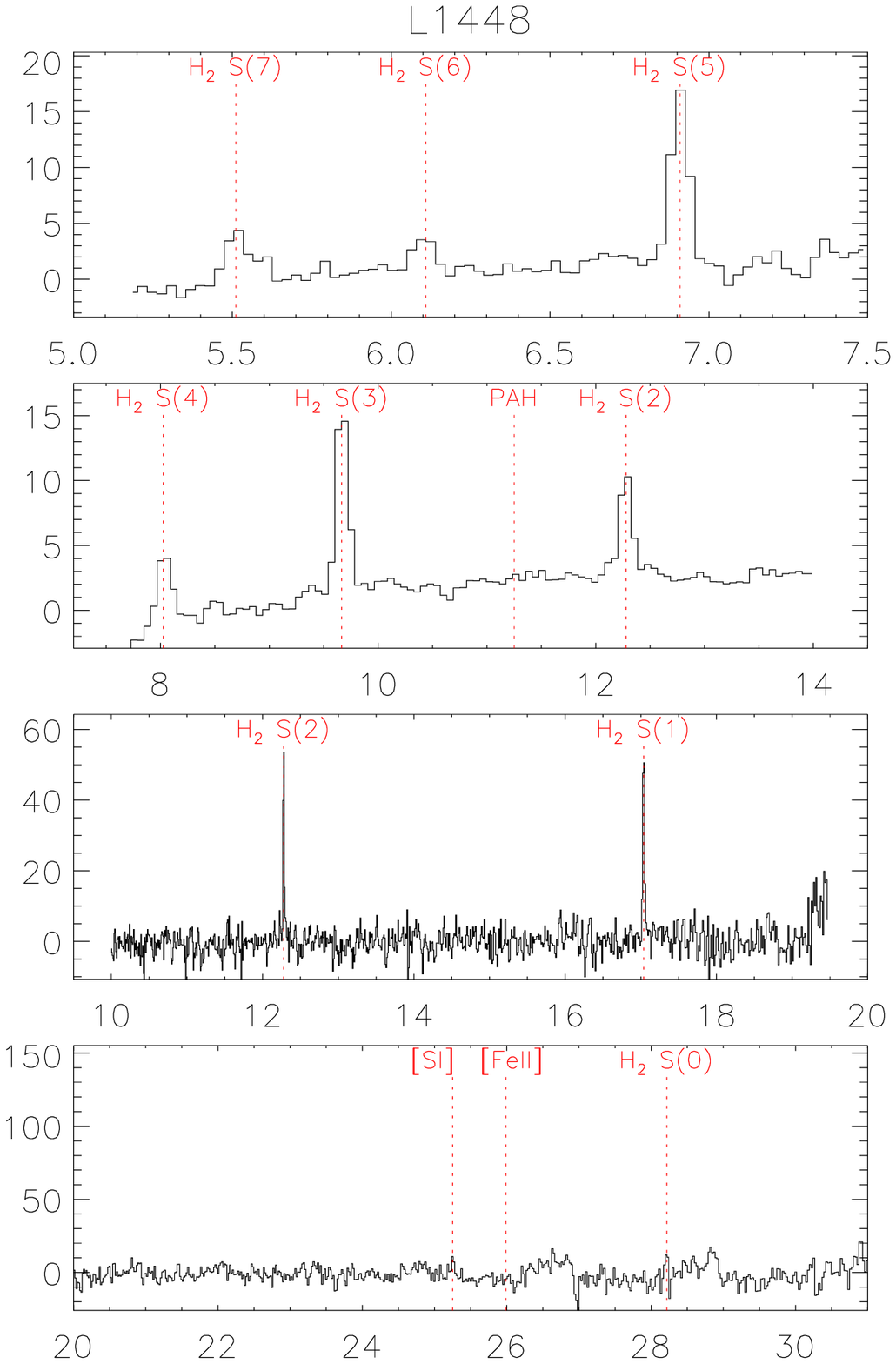}

\noindent{Fig.\ 8 -- same as Fig.\ 1, but for L1448.  
\re{ The beam center position is $\alpha=$ 3h 25m 41.09s, $\delta=$ 30d 42$^{\prime}\,51.8^{\prime\prime}$ (J2000). } }
\end{figure}

\begin{figure}
\includegraphics[scale=1.0,angle=0]{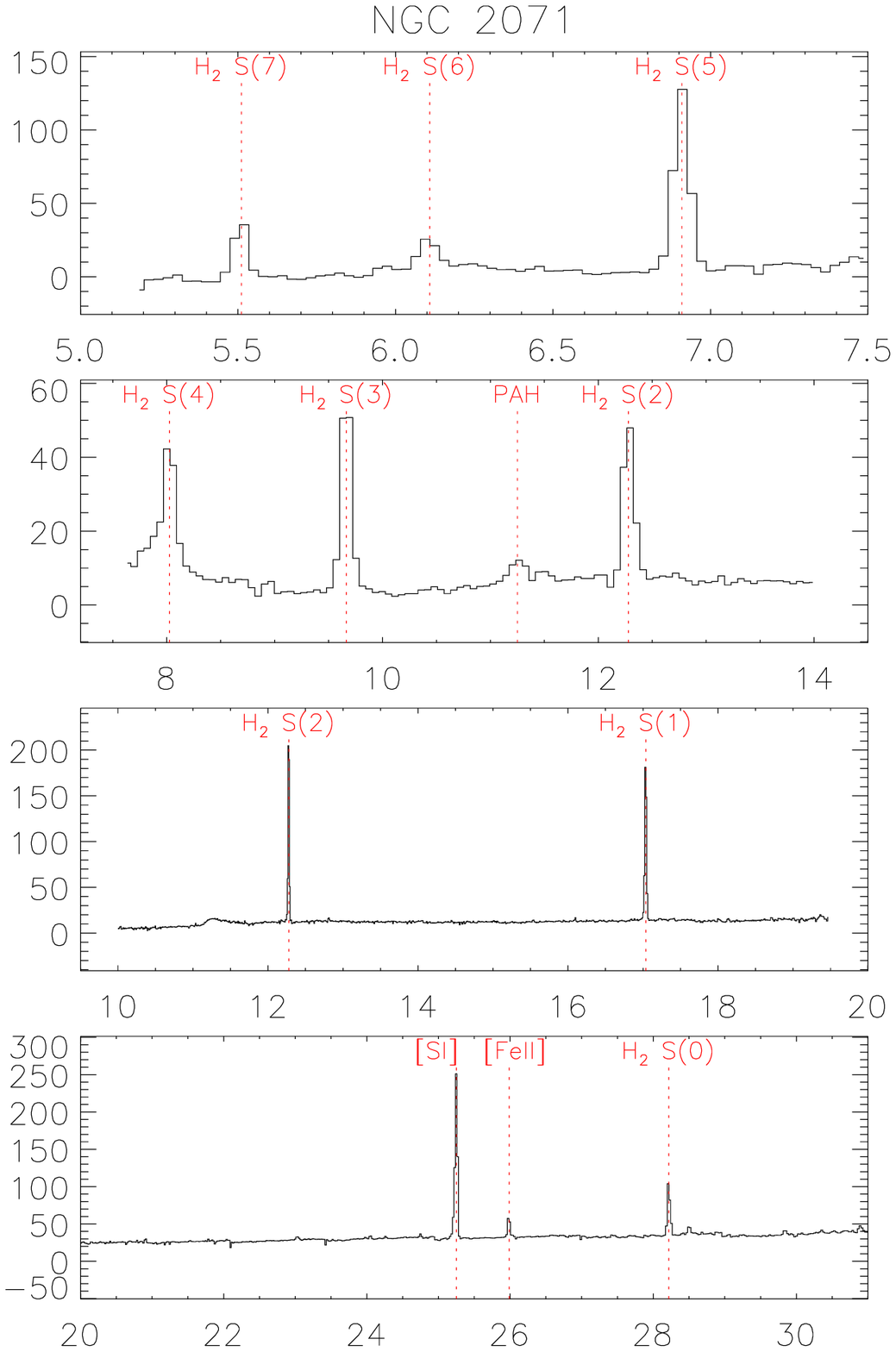}

\noindent{Fig.\ 9 -- same as Fig.\ 6, but for NGC 2071.  
\re{ The beam center position is $\alpha=$ 5h 47m 8.25s, $\delta=$ 0d 22$^{\prime}\,55.3^{\prime\prime}$ (J2000)}.   The weak features at 28.50~$\mu$m and 29.84~$\mu$m are the HD R(3) 
\re{ and H$_2$O $7_{25}-6_{16}$} transitions (discussed previously by Melnick et al.\ 2008) }
\end{figure}

\begin{figure}
\includegraphics[scale=1.0,angle=0]{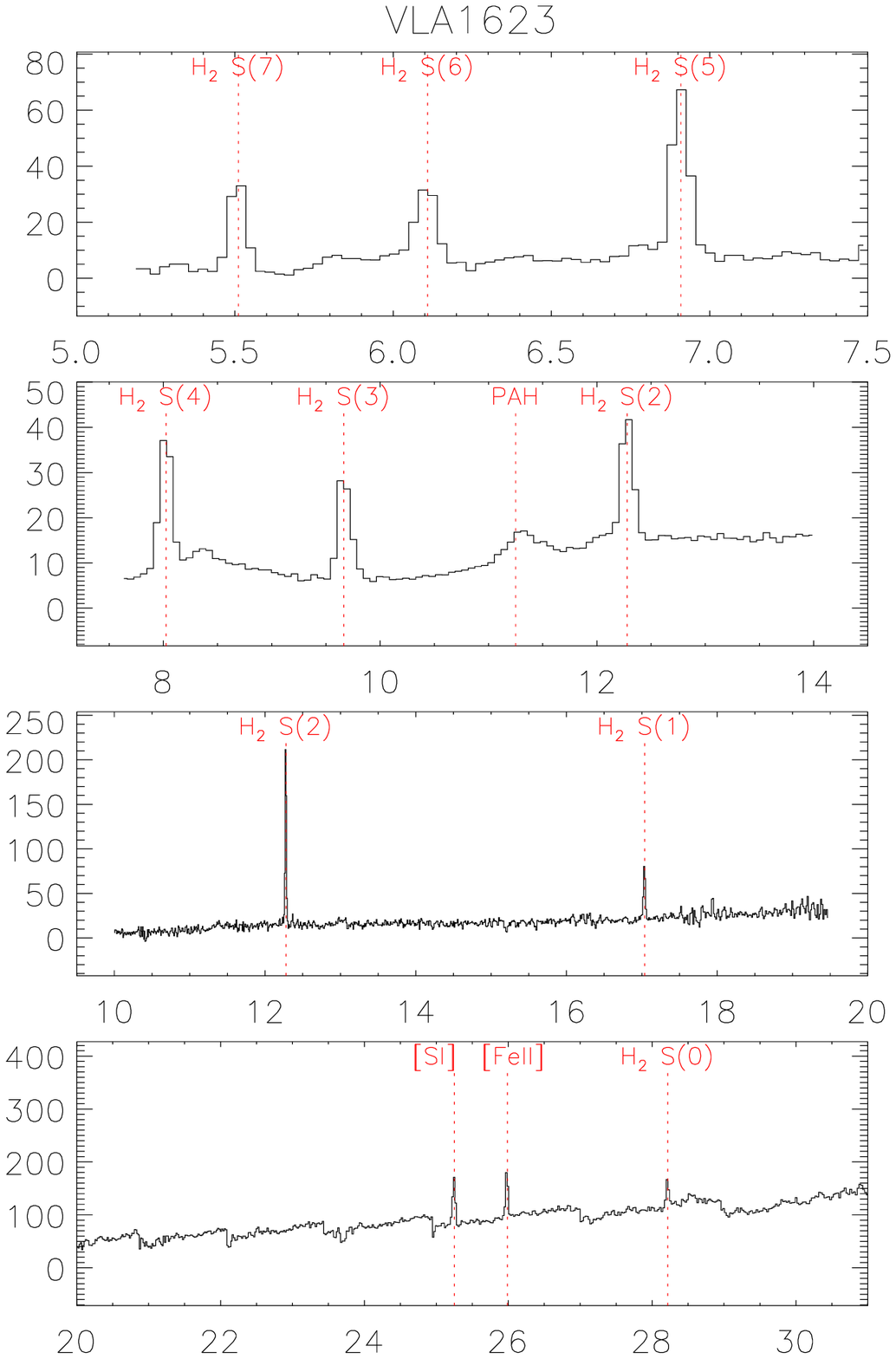}

\noindent{Fig.\ 10 -- same as Fig.\ 6, but for VLA1623.  
\re{ The beam center position is $\alpha=$ 16h 26m 19.08s, $\delta=$ --24d 24$^{\prime}\,47.2^{\prime\prime}$ (J2000) }.  Order mismatches are evident in the LH spectrum (bottom panel).}
\end{figure}

\begin{figure}
\includegraphics[scale=0.85,angle=0]{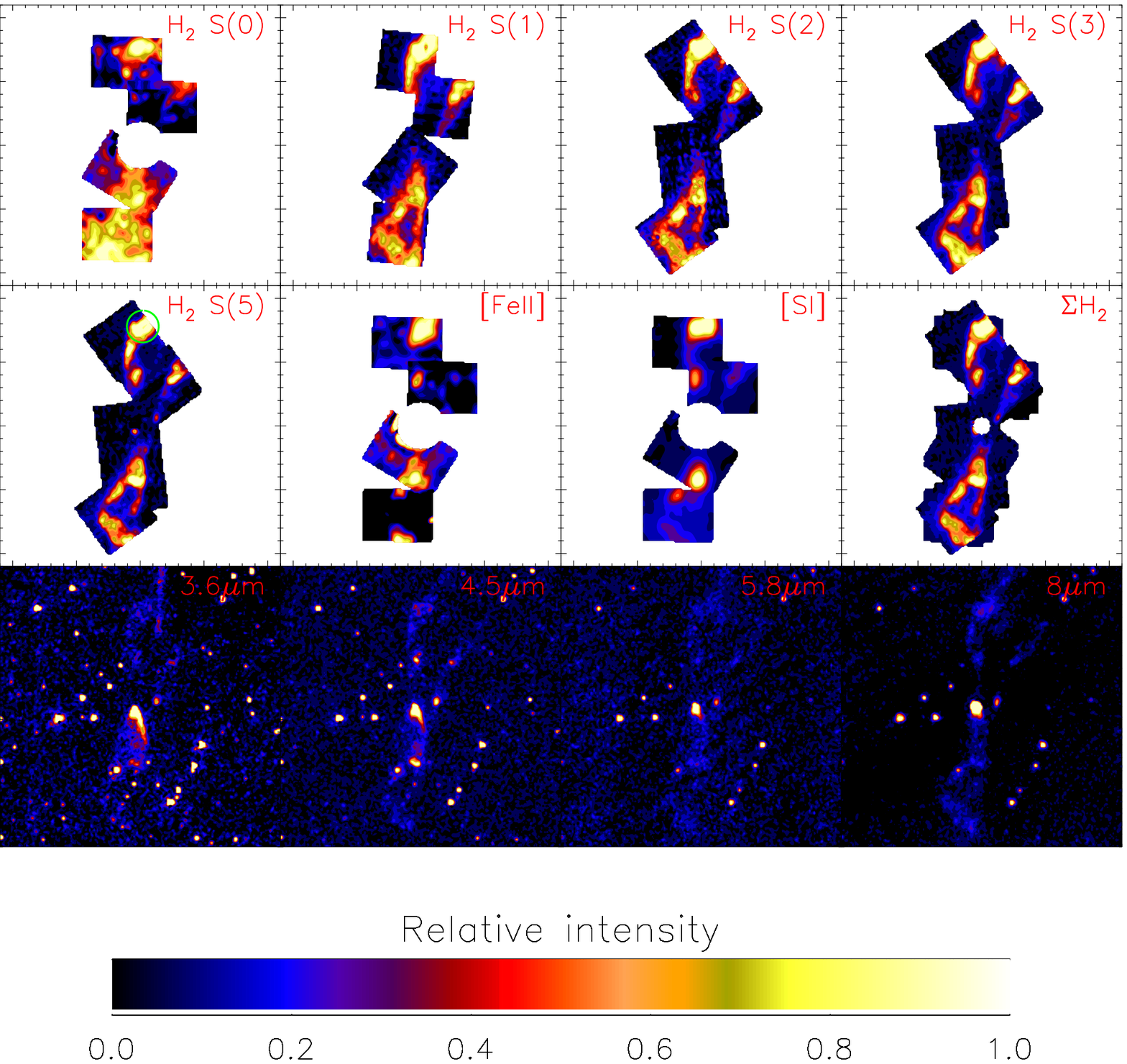}

\noindent{Fig.\ 11 --  Spectral line maps obtained toward BHR71 for the H$_2$ S(0),  
H$_2$ S(1), H$_2$ S(2), H$_2$ S(3), H$_2$ S(5), [FeII] 25.99 $\mu$m and [SI] 25.25 $\mu$m lines, together with the sum of the H$_2$ S(0) -- S(7) emission lines and the archival 3.6 $\mu$m, 4.5 $\mu$m, 5.8 $\mu$m, and 8 $\mu$m IRAC  maps.  Each map is normalized relative to the reference intensities listed in Table 2.  
\re{ The map center position is $\alpha=$ 12h 01m 36.07s, $\delta=$ --65d 08$^{\prime}\,50.5^{\prime\prime}$ (J2000)}, and each small tick interval is 10 arcsec.  
\re{ The green circle indicates the size and position of the 25 arcsec diameter region for which the example spectrum was obtained.  The offset of this position relative to the line center is given in Table 1.} }
\end{figure}

\begin{figure}
\includegraphics[scale=0.85,angle=0]{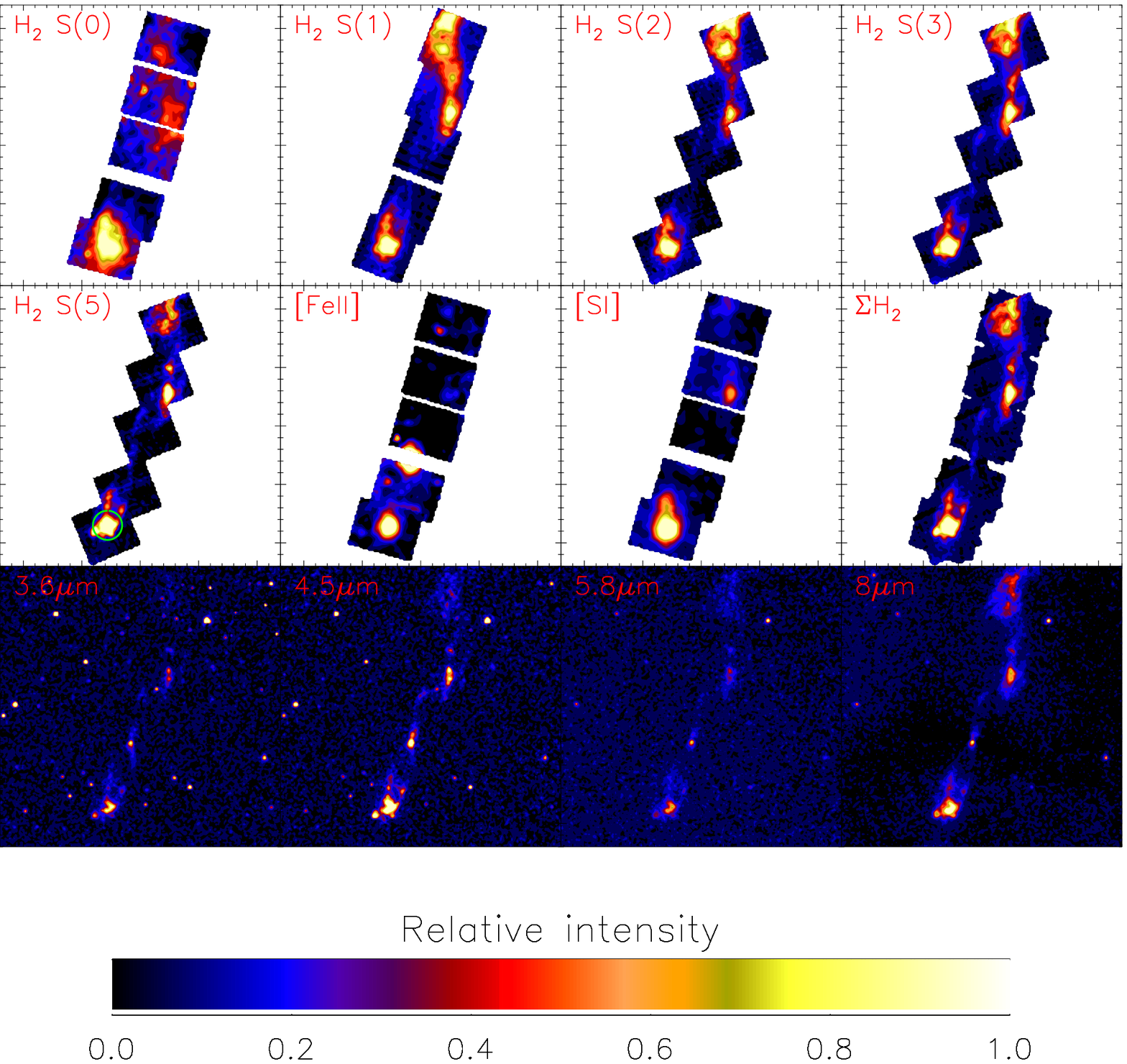}

\noindent{Fig.\ 12 -- same as Fig.\ 11, but for L1157.  
\re{ The map center position is $\alpha=$ 20h 39m 5.11s, $\delta=$ 68d 02$^{\prime}\,43.9^{\prime\prime}$ (J2000).}}
\end{figure}

\begin{figure}
\includegraphics[scale=0.85,angle=0]{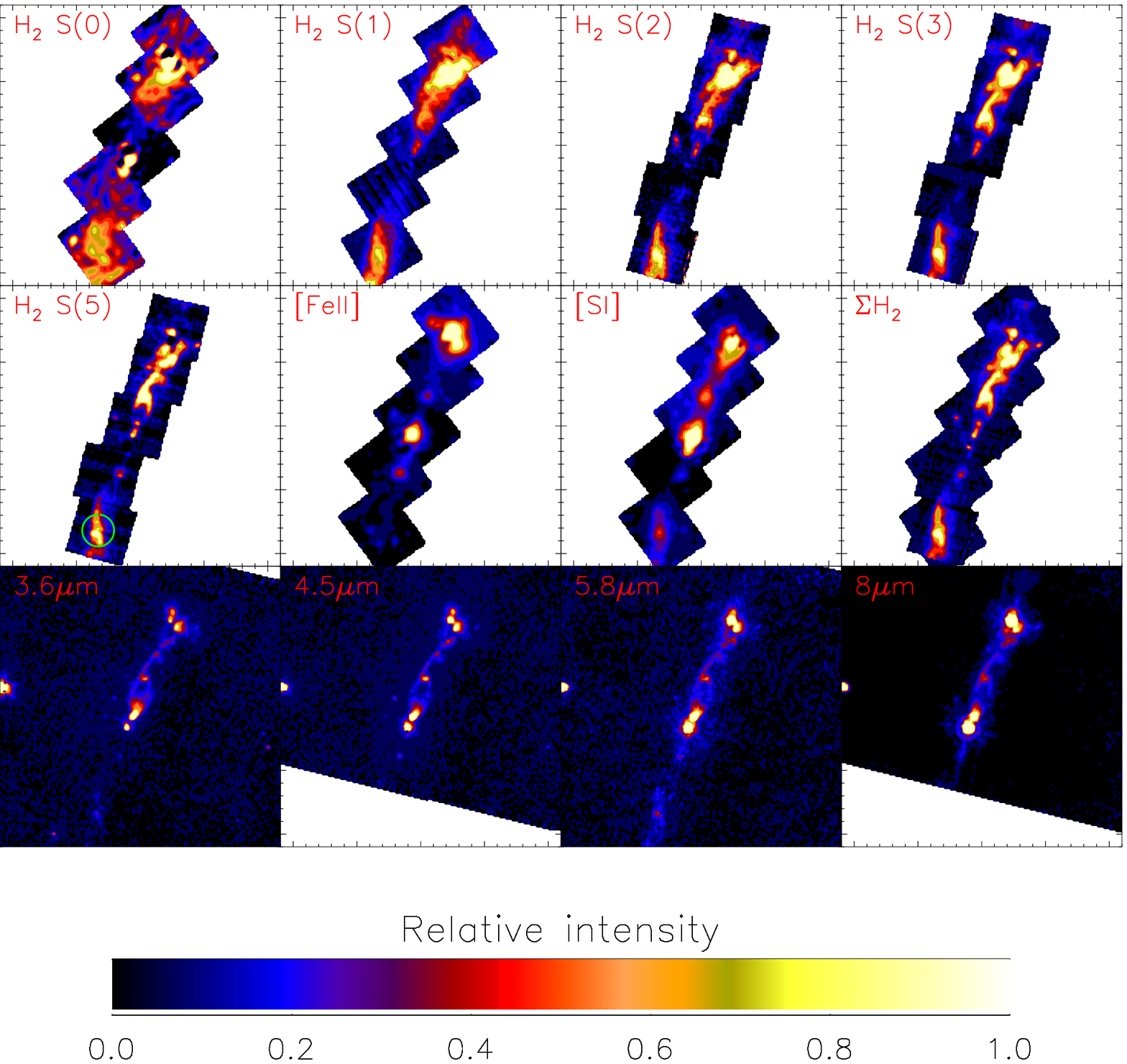}

\noindent{Fig.\ 13 -- same as Fig.\ 11, but for L1448. 
\re{ The map center position is $\alpha=$ 3h 25m 38.45s, $\delta=$ 30d 44$^{\prime}\,13.8^{\prime\prime}$ (J2000).}}
\end{figure}

\begin{figure}
\includegraphics[scale=0.85,angle=0]{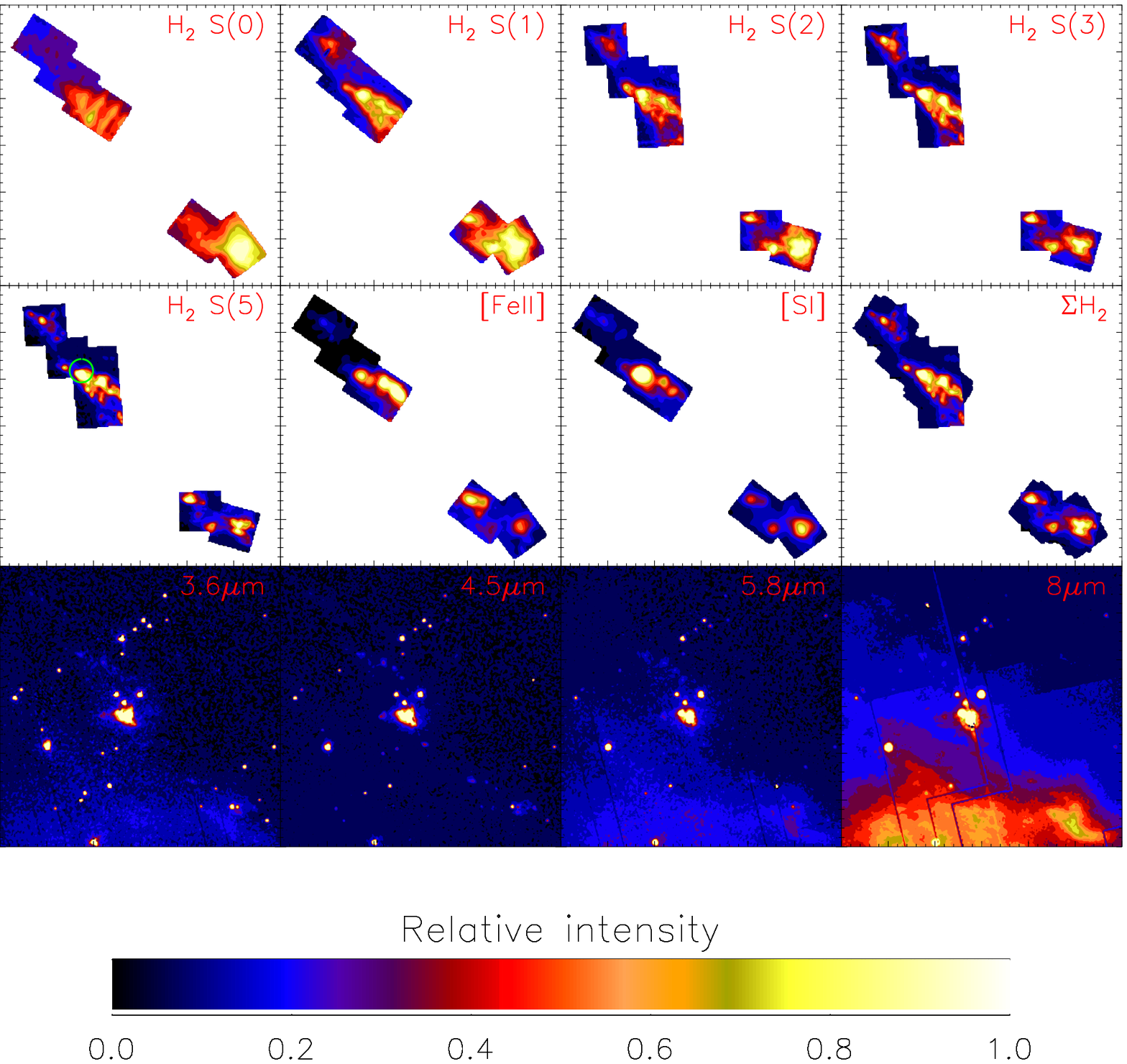}

\noindent{Fig.\ 14 -- same as Fig.\ 11, but for NGC 2071. 
\re{ The map center position is $\alpha=$ 5h 47m 4.05s, $\delta=$ 0d 21$^{\prime}\,56.3^{\prime\prime}$ (J2000).}}
\end{figure}

\begin{figure}
\includegraphics[scale=0.85,angle=0]{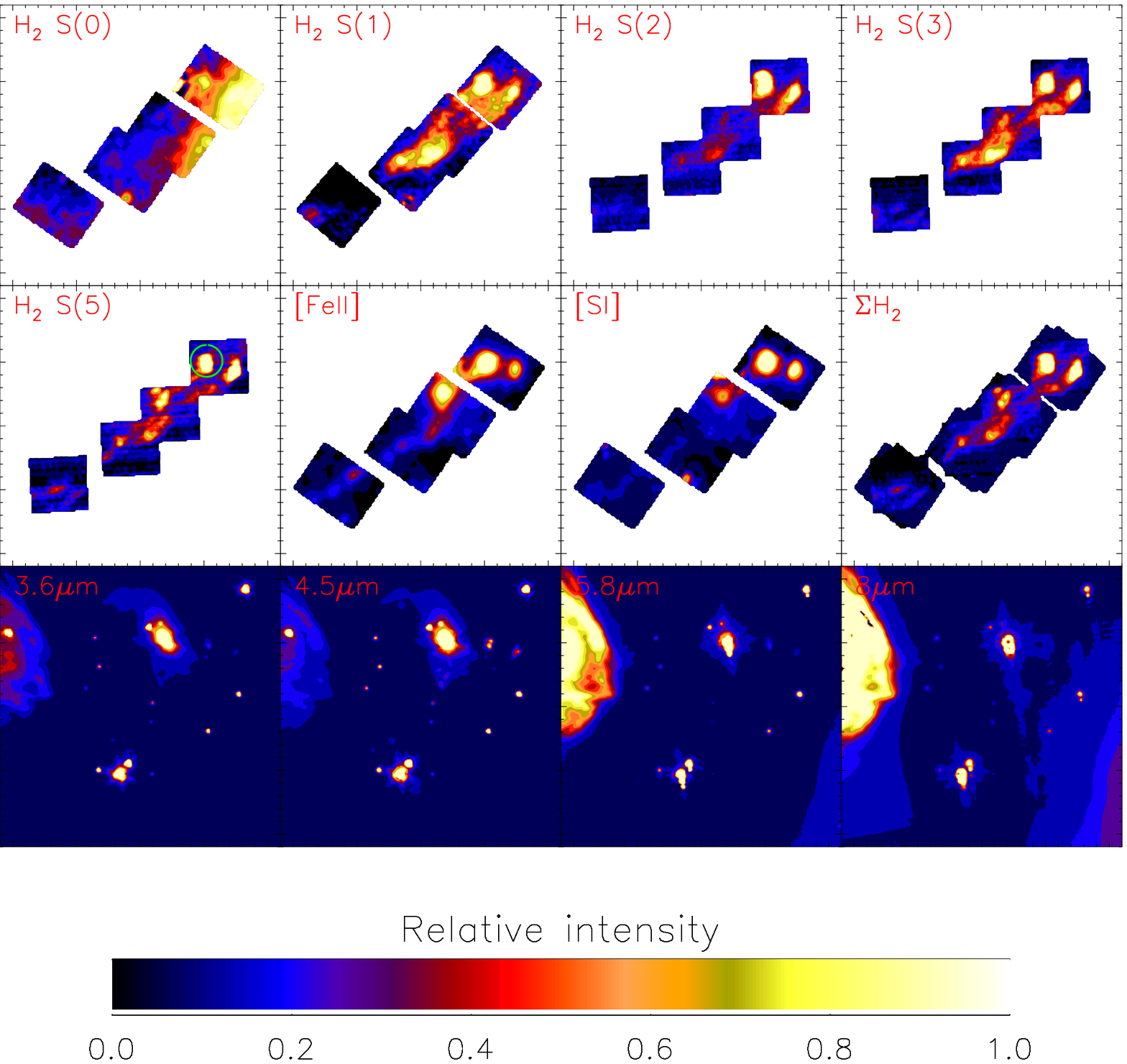}

\noindent{Fig.\ 15 -- same as Fig.\ 11, but for VLA1623. 
\re{ The map center position is $\alpha=$ 16h 26m 22.96s, --24d 23$^{\prime}\,56.2^{\prime\prime}$  (J2000).}}
\end{figure}

\begin{figure}
\includegraphics[scale=0.85,angle=0]{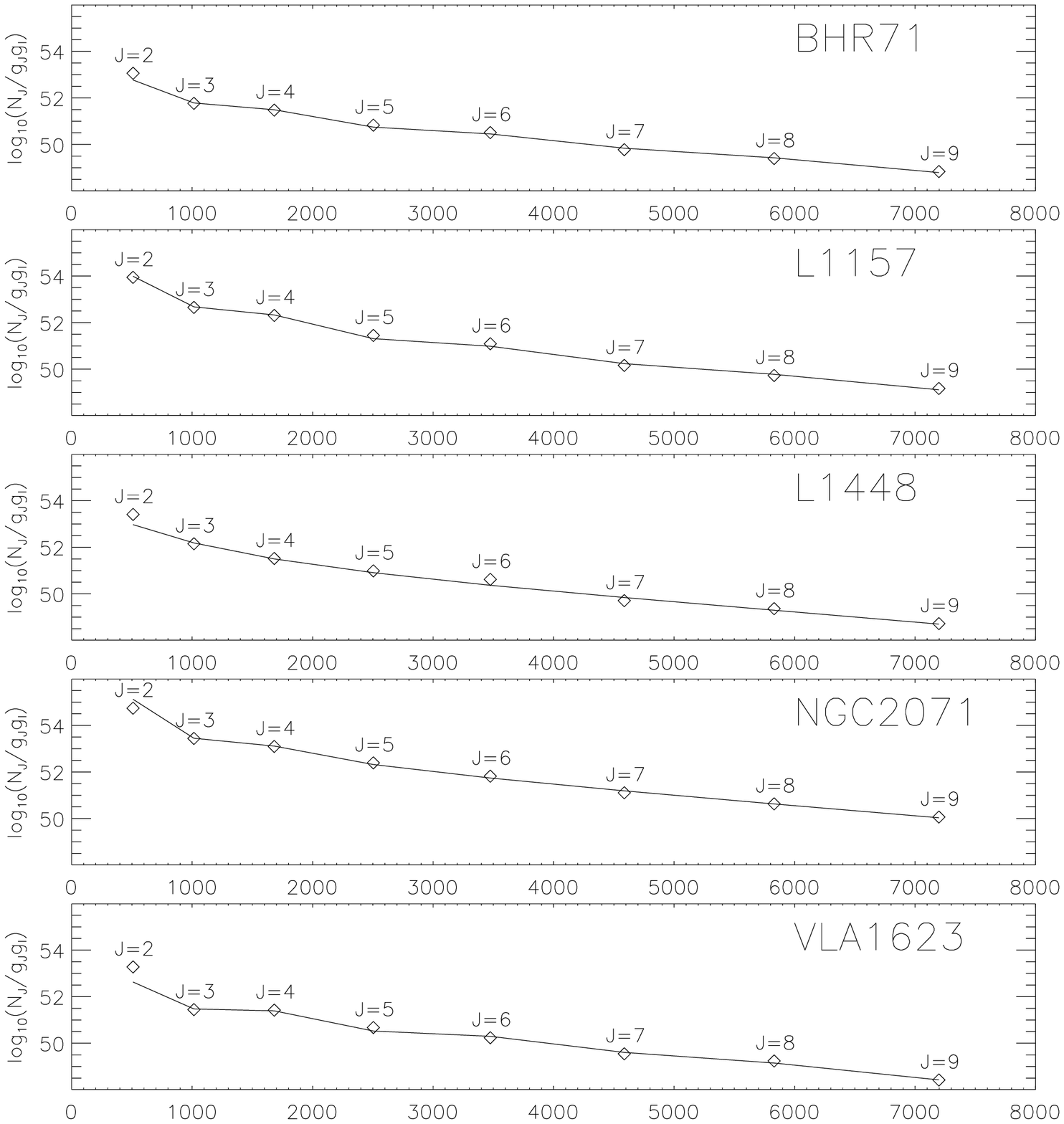}

\noindent{Fig.\ 16 -- Rotational diagrams for each object.  The horizontal axis is the energy of each state, divided by Boltzmann's constant, in units of Kelvin.  The vertical axis is the logarithm of the number of molecules in each state, divided by the product of the degeneracies associated with nuclear spin and with rotation.  Diamonds indicate the observed values, and the solid lines are results obtained from the fitting procedure described in \S4.1}
\end{figure}

\begin{figure}
\includegraphics[scale=0.70]{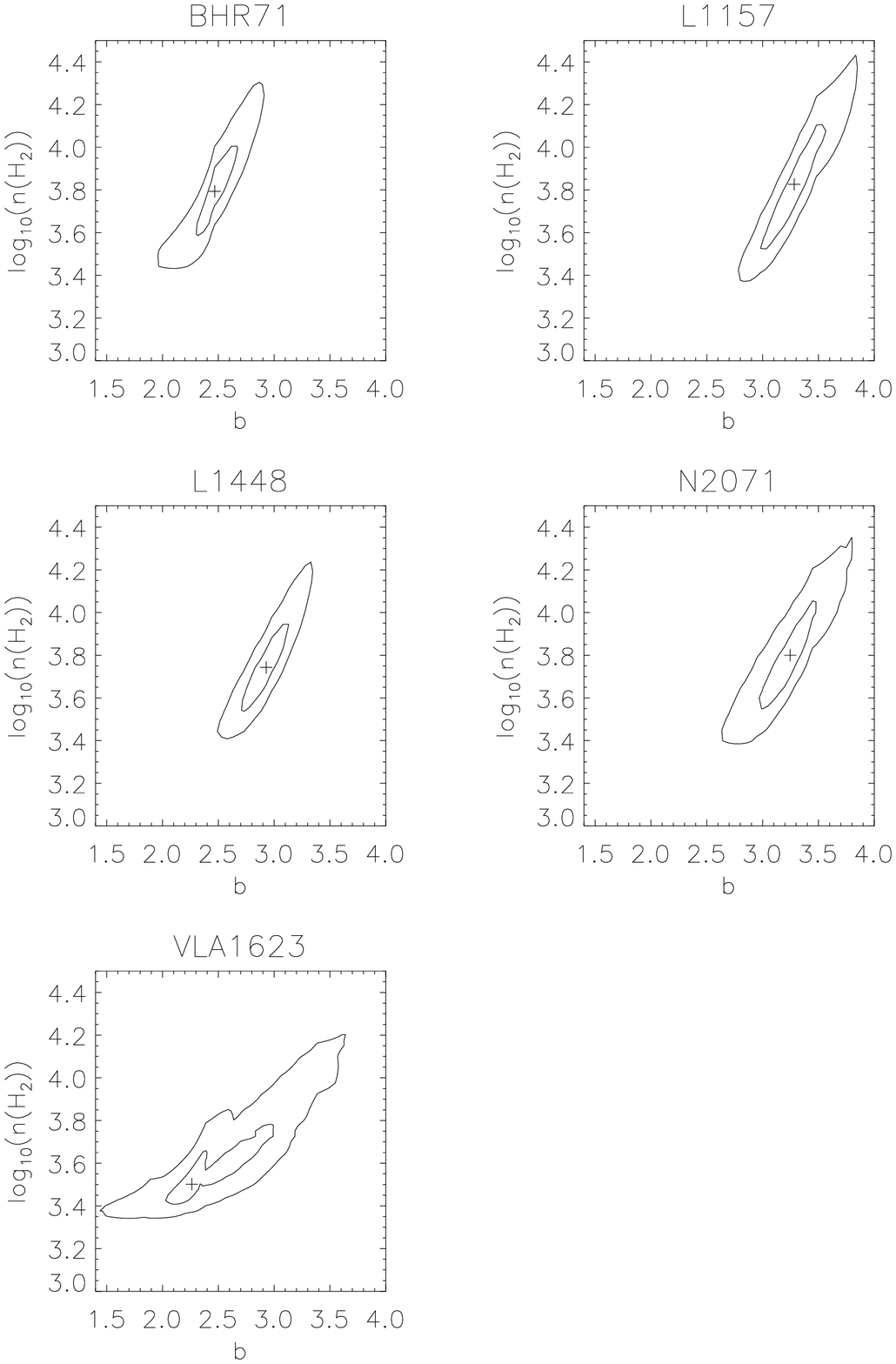}

\noindent{Fig.\ 17 -- Contours of $\chi^2$ in the $b$--$n(\rm H_2)$ plane (see \S4.1).  Crosses indicate the minimum $\chi^2$ (i.e. the best fit), while the two contours represent the 68$\%$ and 95$\%$ confidence limits on each parameter.}
\end{figure}

\end{document}